\documentclass[11pt]{article}
\pdfoutput = 1

\usepackage[utf8]{inputenc}
\usepackage{color,graphicx}
\usepackage{verbatim}
\usepackage{enumerate}
\usepackage{amssymb}
\usepackage{stmaryrd}
\usepackage{physics}
\usepackage{amsfonts}
\usepackage{cite}
\usepackage{array}
\usepackage{setspace}
\usepackage{float}
\usepackage{url}
\usepackage{mathtools}
\usepackage{bbold}
\usepackage{tikz}
\usepackage{graphicx}
\usepackage{wrapfig}
\usepackage{subfigure}
\usepackage[labelfont=bf,font={rm},font={small}]{caption}

\usepackage{tikz}
\usepackage{cite}
\allowdisplaybreaks

\usepackage[margin = 2.2cm]{geometry}
\usepackage[ragged]{footmisc}
\usepackage{hyperref}
\hypersetup{
    colorlinks,
    citecolor=blue,
    filecolor=black,
    linkcolor=blue,
    urlcolor=blue,
    linktocpage=true
}

\def\l{\lambda}

\def\e{{\rm e}}

\newcommand{\be}{\begin{equation}}
\newcommand{\ee}{\end{equation}}
\newcommand{\bea}{\begin{align}}
\newcommand{\eea}{\end{align}}
\newcommand{\bi}{\begin{itemize}}
\newcommand{\ei}{\end{itemize}}

\newcommand{\lr}[1]{\left( #1 \right)}

\def\XXint#1#2#3{{\setbox0=\hbox{$#1{#2#3}{\int}$}
     \vcenter{\hbox{$#2#3$}}\kern-.5\wd0}}

\numberwithin{equation}{section}

\title{Template}

\begin{document}

\thispagestyle{empty}
\begin{center}
\vspace*{.4cm}
  

    {\LARGE \bf 
Extended JT  supergravity and random matrix models:\bigskip\\The power of the string equation
  }
    
    \vspace{0.4in}
    {\bf Clifford V. Johnson 
     and Maciej Kolanowski
    }

\bigskip\bigskip

    {
Physics Department, Broida Hall, University of California, Santa Barbara, CA 93106, USA}
    \vspace{0.1in}
    

    {\tt cliffordjohnson@ucsb.edu},  {\tt mkolanowski@ucsb.edu}
\end{center}

\vspace{0.4in}
\begin{abstract}
\noindent 
A number of  supersymmetric Jackiw-Teitelboim (JT)  gravity theories are  known  to be described (in the Euclidean path integral formulation) by  double-scaled random matrix models. Such matrix models 
can be characterized using a  certain ``string equation''. 
It was shown recently that in extended supergravity, when the number of BPS states scales as ${\rm e}^{S_0}$, where~$S_0$ is the 
extremal entropy, a special ansatz for the leading order solution of the string equation  yields  
the supergravity spectrum.   Somewhat miraculously, the construction showed that 
the functional form of the non-BPS (continuum) sector predicts 
the precise form of the BPS sector,  showing the robustness of  the supergravity/matrix-model correspondence. 
In this paper, we  refine the analysis and show that the  string equation, combined with some simple requirements on  solutions, are powerful 
tools for constraining the spectrum of extended JT supergravity theories. We re-explore the cases of
${\cal N}{=}2$ and  (small) ${\cal N}{=}4$ JT supergravity, and then explore the new cases of spectra from  ${\cal N}{=}3$ and large ${\cal N}{=}4$ JT supergravity (recently derived by Heydeman,  Shi, and Turiaci)  showing that our approach also works naturally for (nearly) all the  models. Based on this success, we conjecture that these new supergravity models also have  matrix model descriptions.
\end{abstract}
\pagebreak
\setcounter{page}{1}
\tableofcontents

\newpage

\section{Introduction}
\label{sec:introduction}

\subsection{Opening Remarks} 

Since the striking discovery of ref.~\cite{Saad:2019lba} that Jackiw-Teitelboim (JT) gravity~\cite{Jackiw:1984je,Teitelboim:1983ux}, when formulated using the Euclidean gravitational path integral (GPI), has a description as a double scaled~\cite{Brezin:1990rb,Douglas:1990dd,Gross:1990vs, Gross:1990aw} random matrix model (RMM),  there has been steady progress in generalizing the duality to models of JT gravity with various types of supersymmetry. Among the reasons why this is particularly interesting is the fact that many such JT supergravity models can arise in the low energy near-horizon geometries of supersymmetric higher dimensional black holes (see {\it e.g.,} refs.~\cite{Heydeman:2020hhw,Boruch:2022tno}) for which there is good understanding of aspects of their microscopic dynamics from complementary approaches such as direct D-brane microstate counting~\cite{Strominger:1996sh}, and the AdS/CFT correspondence~\cite{Maldacena:1997re}.

In fact, the cases of JT supergravity with  extended supersymmetry have only recently been understood in matrix model terms, although the picture is evolving swiftly. The case of ${\cal N}{=}2$ was analyzed in detail in ref.~\cite{Turiaci:2023jfa}, with some discussion of aspects of the ${\cal N}{=}4$ case presented there as well. Ref.~\cite{Johnson:2023ofr} presented a formulation of the matrix model for the ${\cal N}{=}2$  case in terms of a combination of multicritical models, allowing for access to non-perturbative (in genus) physics as well. Furthermore, in ref.~\cite{Johnson:2024tgg}, an analysis in terms of multicritical models  was presented for the  ${\cal N}{=}4$ case as well. That same paper argued that  the important presence of a gap in the extended JT supergravity spectrum (and hence the low energy black hole spectrum) when there is a large number of BPS states is a robust feature that follows from  a random matrix description.  The key point is that random matrix models that capture the gravitational physics  with large numbers of BPS states are very naturally constructed from multicritical generalizations~\cite{Morris:1990bw,Dalley:1992qg} of models of random matrices of Wishart form $M^\dagger M$ (where~$M$ is a complex matrix)  in  the double-scaling   limit. Within such models, eigenvalue repulsion from the large number of degenerate states pushes the rest of the spectrum away, forming the gap that appears in supergravity.

The multicritical descriptions of interest are captured (both perturbatively and non-perturbatively) in terms of a ``string equation'', a non-linear ordinary differential equation (ODE)  of a very particular form. Very interestingly, the ansatz found in ref.~\cite{Johnson:2023ofr} that  defines the appropriate class of string equation solutions shows that not only is there a connection between the BPS and non-BPS sectors through the dynamical generation of a gap, but that  trying to construct a multicritical description of the non-BPS part of the spectrum {\it fails} unless the BPS sector is of {\it precisely} the (non-trivial) functional form given by the supergravity analysis!

This connection is remarkable, and deserves to be better understood. That is the purpose of the present paper.  The understanding  reported here represents the result of analyzing and refining the approach of ref.~\cite{Johnson:2023ofr}, placing it on firmer ground. The connection between the BPS and non-BPS sectors follows from extending to the complex $E$ plane the leading density $\rho_0(E)$ that results from  the special form  of  the string equation, whereupon some simple assumptions combined with analyticity connect the BPS and non-BPS sectors in a very precise way. 

Another key observation to be made is that the tight connection enforced by the string equation can be turned around to be predictive. Two examples of this are as follows: 

\begin{enumerate}[{\bf (1)}] 

 \item That a miracle seemed to occur in order for ${\cal N}{=}2$ and  (small) ${\cal N}{=}4$ to work out---an infinite number of negative powers of $E$ needed to cancel (see Section~\ref{sec:the-miracle} for a review)---led to a conjecture\footnote{\label{fn:pattern}CVJ, in unpublished notes from Winter 2024: A strong motivation for the conjecture came from the alternating pattern of the occurrence in $\rho_0(E)$, in the numerator, of (schematically)  $\sinh(2\pi\sqrt{E})$ and $\cosh(2\pi\sqrt{E})$ as ${\cal N}$ is increased, with~${\cal N}/{2}$ powers of~$E$ in the denominator. There was an obvious gap in the pattern corresponding to ${\cal N}{=}3$. (See Section~\ref{sec:Closing-Remarks} for further discussion of this.) A naive count of the available fermionic and bosonic zero modes seemed to bolster the idea. Further  circumstantial evidence came from the apparent existence of an ${\cal N}{=}3$ Schwarzian derivative in 2D superconformal field theory~\cite{Schwimmer:1986mf,Chang:1987qj}.} of the existence of ${\cal N}{=}3$ JT supergravity models as another context where the same miracle had a chance to occur. This has been beautifully confirmed in detail by the new ${\cal N}{=}3$ models presented recently in ref.~\cite{Heydeman:2025vcc}! We analyze one of them in Section~\ref{sec:Neq3-JT-supergravity}.

  \item  For the new ``large'' ${\cal N}{=}4$ JT supergravity, the  non-BPS part of the spectral density, computed in ref.~\cite{Heydeman:2025vcc}, does not fit with the allowed form that our analysis shows is required by the string equation for a multicritical model. (See Section~\ref{sec:large-Neq4-JT-supergravity}.) Instead, we find that it is the sum of two parts that each separately {\it does} fit. Furthermore, each part separately  appears as different terms in the derivation of the partition function in  ref.~\cite{Heydeman:2025vcc}. Treating it as our input, we find again that the BPS physics can be recovered from knowledge of these non-BPS sectors. This is even more surprising than for the simpler models because now there is an infinite number of short multiplets with different BPS energies and different degeneracies. We recover all of these data, together with a non-trivial condition on  what spins there are any non-zero BPS states. This success strongly suggests that our separate treatment of each sector is justified {\it i.e.,} the two sectors are statistically independent from each other, and each separately have a random matrix model description, while their sum does not. It would be interesting to test out this prediction by a gravity computation. Presumably the wormhole/cylinder diagram with boundaries coming from different sectors would vanish.
\end{enumerate}

Beyond the strong constraints on perturbative physics afforded by the string equation, there are powerful non-perturbative conclusions that follow as well. For some values of the parameters, solutions that appear to be fine perturbatively (to all orders)  simply cannot have emerged as the limit of any sort of non-perturbative solution of the equation, suggesting that they are ruled out entirely. (See Section~\ref{sec:non-perturbative-remarks}.)

It is therefore evident that the string-equation-based techniques and framework presented here is a powerful toolbox for gaining better understanding of extended JT supergravity and the many higher dimensional black hole systems from which they naturally emerge.

\subsection{Outline}
\label{sec:outline}
Since  string equations and the role they play are less familiar to some readers,  it is perhaps prudent to review and refine some of the intuition behind them, as well as briefly review the multicritical matrix model approach. This is the purpose of the next  Section (\ref{sec:string-equations-general}). Section~\ref{sec:JT-supergravity-multicritical} then reviews how JT supergravity models are built as multicritical models, and in particular Subsection~\ref{sec:the-miracle} discusses the miraculous precise connection between the BPS and non-BPS sectors that  the string equation (with the ansatz of ref.~\cite{Johnson:2023ofr})
requires.

Having completed review,  Section~\ref{sec:new-approach} begins with reflection on the strengths and limitations of the ansatz, and then develops a new, refined approach to building a multicritical matrix representation of  extended JT supergravity. A key element of this approach is an energy expansion about $E_0$, which is better behaved. It removes the necessity to handle an infinite number of inverse powers of $E$. Instead, the analysis boils down to how the solution for $u_0(x)$ translates into analytic properties of the spectral density $\rho_0(E)$ on the complex $E$-plane, yielding a clearer picture of what underlies the ``miracle'' of the earlier method of ref.~\cite{Johnson:2023ofr}. 

The approach presented is quite general, and so in Sections~\ref{sec:Neq2-JT-supergravity} and~\ref{sec:small-Neq4-JT-supergravity} respectively, the ${\cal N}{=}2$ and small ${\cal N}{=}4$ JT supergravity are re-done from the new perspective.
This is followed by an analysis of the new cases of large ${\cal N}{=}4$ (in Section~\ref{sec:large-Neq4-JT-supergravity}),   and ${\cal N}{=}3$ (in Section~\ref{sec:Neq3-JT-supergravity}). 
Section~\ref{sec:exploration} discusses non-perturbative aspects of the models, and some general features of note. We end with some closing remarks in Section~\ref{sec:Closing-Remarks}.

\subsection{The Language of String Equations}
\label{sec:string-equations-general}
\subsubsection{Natural  Measures}
{In the broad context of JT gravity (and the spectral diagnostic of low energy black hole dynamics that it provides\cite{Turiaci:2023wrh}), the core role of the random matrix model is as an ensemble of Hamiltonians $H$ that are candidate holographic duals of the (effective) 2D gravity system.  As such, for JT {\it super}gravity they 
should have a positive spectrum of eigenvalues, following from  the fact that~$H$ is written in terms of a supercharge $Q$ as $H{=}Q^2$. 

As discussed in ref.~\cite{Stanford:2019vob} (see also ref.~\cite{Turiaci:2023jfa}), how one proceeds depends upon whether a  $(-1)^F$ symmetry is preserved or not. In a representation of $(-1)^F$  as block diagonal, preserving it means that~$Q$ should anticommute with it, and so  can be written as:
\begin{equation}
\label{eq:0A-Q-matrix}
    Q=\begin{pmatrix}
    0 & M \\ M^\dagger & 0
    \end{pmatrix}\ ,
\end{equation} 
where $M$ is a complex matrix. In general, it will be thought of as an $(N{+}\Gamma){\times}N$ rectangular matrix, and conventions are chosen here such that the matrix $M^\dagger M$ is an $(N{+}\Gamma){\times}(N{+}\Gamma)$ square matrix with~$\Gamma$ repeated zero eigenvalues. This is the ``type 0A'' variant of model. Without the need to anticommute with a $(-1)^F$ the only constraint on  $Q$ is that it is  Hermitian, of size $N{\times}N$. Such models constitute the ``type 0B'' variant. 
The partition functions for these matrix models are simply of the form:
\begin{equation}
\label{eq:partition-functions-A-B}
    {\cal Z}_{\rm A} = \int\! dM \e^{-N{\rm Tr}[V(M^\dagger M)]}\ ,\quad\mbox{\rm or}\quad  {\cal Z}_{\rm B} = \int \!dQ\,\, \e^{-N{\rm Tr}[V(Q)]}\ ,
\end{equation}
with a simple polynomial potential $V$ with quadratic and higher terms. These are {\it not} to be confused with the gravity partition function $Z(\beta)$ to be discussed later. Instead, that is the expectation value of a specific combination of matrices defining a loop observable of length $\beta$, where $\beta$ is the period of Euclidean time.

In the above, the trace means that the physics boils down to an effective model   of the eigenvalues, after using gauge invariance  to write~$M$ or~$Q$ in the form $U\Lambda W^\dagger$ where $U$ and $W$ are diagonalizing matrices and $\Lambda$ is diagonal. There is a Jacobian, $J$, for changing from the matrices to their eigenvalues. In both variants, $J$ contains the square of the Vandermonde determinant $\Delta(w){\equiv}\prod_{i<j}(w_i-w_j)$, where~$w_i$ is the appropriate eigenvalue variable.  After taking the logarithm and exponentiating, this can be taken to be part of the potential, adding a logarithmic repulsion between all eigenvalues. It is  a ``Dyson gas"  or ``log gas'' model~\cite{ForresterBook,Meh2004}.

The natural eigenvalues, $\lambda_i$, of the problem for type~0A are those of $M^\dagger M$ or $M M^\dagger$. The difference is that the former system has  $\Gamma$ zero eigenvalues while the latter does not. Henceforth we will write formulae with the $M^\dagger M$ system in mind, since we are interested in systems with BPS states, which, as mentioned above,  $\Gamma$ will count.
The  $\lambda_i$  naturally live on~$\mathbb{R}^+$, which can be interpreted as a ``wall'' in the problem that prevents eigenvalues (particles of the gas model) from going to negative values (positions). There is an extra factor $\prod_i\lambda_i^\Gamma$ in the Jacobian that (after exponentiation) gives the repulsive contribution to the potential: ${\cal V}_\Gamma{\equiv}{-}\Gamma\sum_i\log \lambda_i$. Written in terms of the eigenvalues $y_i$ of~$M$ (where~$y^2_i{=}\lambda_i$) the action for the 0A system is (after dividing out a factor of the volume of the unitary group):
\begin{equation}
\label{eq:dyson-gas-0A}
 {\cal Z}_{\rm A} = \int_{0}^\infty\! \prod_i^N 
 d\lambda_i\prod_{i<j}(\lambda_i - \lambda_j)^2\prod_i^N \lambda_i^{\Gamma}\,\,\e^{-N\sum_i^NV(\lambda_i)}
  = \int_{-\infty}^\infty\! \prod_i^N 
 dy_i\prod_{i<j}(y_i^2 - y_j^2)^2\prod_i^N y_i^{2\Gamma+1}\,\,\e^{-N\sum_i^NV(y_i^2)}\ ,
\end{equation}
and from the second form, it is referred to as an $(\boldsymbol{\alpha},\boldsymbol{\beta}){=}(2\Gamma{+}1,2)$
ensemble in the Altland-Zirnbauer classification\cite{Altland:1997zz}. The number $\boldsymbol{\beta}$ refers to the power of Vandermonde detertminant $\Delta$. While $\Gamma$ was introduced naturally as an integer here, it can be take more general values. For example, the cases where~$\Gamma$ is half integer are also natural here.\footnote{They  can be obtained (at least locally) from other matrix ensembles (such as unitary matrix models), resulting in this same effective Dyson gas model~(\ref{eq:dyson-gas-0A}) after a change of variables~\cite{Dalley:1992br}.} 

For this paper, the interest will be in models where, on taking the large $N$ limit, $\Gamma$ is also scaled with~$N$.  The focus will also be on variants of JT gravity, where there is a parameter $S_0$ which (often) has  the interpretation  the $T{=}0$ ``extremal entropy'' of a higher dimensional black hole for which the (2D) JT gravity system has emerged as the effective ``throat" description of the low temperature near-horizon dynamics. In a quantum gravity description of the physics,  the dimension of the black hole's Hilbert space is of order $N{=}\e^{S_0}$  (at least semi-classically), and large $N$ random matrix models have emerged~\cite{Saad:2019lba,Stanford:2019vob} as good descriptions of JT gravity and various generalizations thereof. Hence, having $\Gamma{\sim} \e^{S_0}$ makes the 0A-type matrix models under discussion here  particularly appropriate for studying {\it supersymmetric} black hole systems systems that have throats, and additionally sectors with some number of  BPS states scaling as $\e^{S_0}$. 

Following this line of investigation, ref.~\cite{Johnson:2024tgg} argued that the presence of  gaps in the spectrum uncovered in the supergravity analysis of such black holes has a natural explanation within these matrix models: The  term representing a repulsion from the origin, discussed above, $V_\Gamma{\equiv}{-}\Gamma\sum_i\log \lambda_i$ is normally subleading in the large $N$ limit, {\it i.e., } 
using a continuous variable $X{=}i/N$ in the limit, ${\cal V}_\Gamma\,{\to}\,{-}\frac{\Gamma}{N}\int_0^1\log[ \lambda(X)]\, dX$, which is subleading compared to the physics controlled by $V(\lambda )$ in this limit. If however $\Gamma$ scales with $N$, {\it i.e.,} $\Gamma{=}{\widetilde\Gamma}N$, then the resulting repulstive potential ${\cal V}_\Gamma\,{=}\,{-}{\widetilde\Gamma}\int_0^1\log [ \lambda(X)]\, dX$ is of the same order as the leading physics. This natural (and quite generic) matrix model mechanism therefore is responsible for a (super)gravity gap in systems with a large degeneracy of states. The systems to be discussed at length in this paper are examples of this kind of system, and we will argue that they all have matrix model descriptions, and demonstrate (as begun in ref.~\cite{Johnson:2024tgg}) that this is not merely qualitiative: The precise features of the supergravity spectrum, the non-BPS sector, the gap, the BPS degeneracy (both its location and multiplicity) are all tightly constrained in relation to each other in order to have a consistent random matrix model description! This will all emerge from the properties of the leading part of a particular ``string equation''~(\ref{eq:big-string-equation}) (to be reviewed below) that describes the matrix model.

\def\ty{{\tilde y}}

Of course, the type~0B choice (see below~(\ref{eq:0A-Q-matrix})) should not be forgotten, and although much won't be said about it in this paper, it should naturally enter the JT extended supergravity story too.  In this case, the natural eigenvalues (denoted $\ty$) are simply those of the Hermitian~$Q$, and valued on $\mathbb{R}$. The eigenvalue model is, in the  simplest case:
\begin{equation}
\label{eq:dyson-gas-0B}
    {\cal Z}_{\rm B} = \int_{-\infty}^\infty\! \prod_i^N 
 d\ty_i\prod_{i<j}(\ty_i - \ty_j)^2\,\, \e^{-N\sum_i^N{\widetilde V}(\ty_i)}\ ,
\end{equation}
In this case, the  $\lambda_i{=}\ty_i^2$, the square of $Q$'s eigenvalues, give the naturally positive eigenvalues of $H$. The measure/Jacobian combination above gives a $\boldsymbol{\beta}{=}2$ Dyson ensemble. 
Absent here is a natural parameter corresponding to the count (scaling with $N$) of a BPS degeneracy (analogous to $\widetilde\Gamma$ of the 0A formulation above). While there are candidate such analogues\footnote{For example, there are non-perturbative maps~\cite{Morris:1990bw,Dalley:1992br,Klebanov:2003km} between the solutions of the string equations for the 0A system and the solutions of the string equations for the 0B system (and generalizations thereof) that turn $\Gamma$ into a parameter on the 0B side. See ref.~\cite{johnson:2025vyz} for a recent discussion and application. More generally, $Q$ can be constrained (presumably to have degeneracies) by coupling it to a fixed external matrix of an appropriate type.}, they will not be explored in this paper. Some comments about the possible need for 0B realizations of ${\cal N}{=}3$ models will be made in  Section~\ref{sec:Neq3-JT-supergravity}.

\subsubsection{Multicriticality}
\label{sec:multicriticality}
The next key idea is that of multicriticality. With  just Gaussian random matrix models (quadratic potential $V(\lambda)$~(or~${\widetilde V}({\tilde y})$)), while the resulting spectral problems already capture interesting physics, they are limited in the kind of {\it gravitational} physics they can capture. While many topological aspects can be modeled, 
 if  the observables include physics sensitive to the areas of 2D surfaces, Gaussian models  will not suffice.  Modelling gravitational systems where surfaces with non-zero area survive in the continuum limit is made possible with higher order potentials.

 At large $N$, the eigenvalues can be treated as continuous and the leading density is a  function $\rho_0(\lambda)$ that (generically) has endpoints. Much can be said about the type of physics that will be accessible by classifying the behaviour of $\rho_0(\lambda)$. Since for given $\boldsymbol{\beta}$,  the local behaviour of the interior of  one random matrix  model's spectral density looks much like the other, the distinguishing features are to be found at the endpoints/edges. The 0B model generically has the classic ``soft edge'' behaviour, while 0A is interesting for its ``hard edge'' behaviour, where at one end the eigenvalues bump into the aforementioned wall.  Allowing the potentials to have more than quadratic behaviour allows for more, however. For a start, in the 0A system, $\rho_0(\lambda)$ can be tuned to pull away from the wall entirely and have (leading) soft edge behaviour at both ends. Furthermore, the 0B system can produce ``bumping'' behaviour of a related (but different) sort by having two components to the spectral density (in $Q$ eigenvalues) bump into each other, giving new behaviour when they merge.\footnote{The classic Gross-Witten-Wadia phase transition~\cite{Gross:1980he,Wadia:1980cp} is the prototype of this.} This comes from tuning  an even potential that produces a symmetric double-well problem in the Hermitian matrix $Q$.

The universality or critical behaviour that is of interest is characterized by the rate of fall-off of  $\rho_0(\lambda)$ toward zero. Generically for the classic Wigner semi-circle type behaviour it is a square root $\rho_0{\sim}(\lambda-\lambda_-)^\frac12$.  It is multiplied by an additional $\lambda^{-1}$ in the Wishart (0A) case and $\lambda_-$ is at the wall (the  local piece of the classic Marchenko-Pastur-type 
  behaviour\cite{Pastur:1967zca}).  ``Multicriticality''~\cite{Kazakov:1989bc}, indexed by an integer $k$, results from tuning the potential such that there are $k{-}1$ additional zeros at $\lambda_-$. Tuning parameters in the potential to attain  such  multicritical behaviour while also sending $N\to\infty$ allows for different kinds of models of gravity to be obtained in the ``double scaling limit''\cite{Douglas:1990ve,Brezin:1990rb,Gross:1990vs,Gross:1990aw}.
To understand where this comes from it is useful to think of another role that these random matrix models have. Treating them as toy field theories, their Feynman graphs ('t Hooft ribbon graphs) give (after a dual decoration) tesselations of the  sum over surfaces (appropriately ordered by topology - there is a factor $N^{-\chi}$ for a surface where $\chi$ is the Euler number) required to yield the 2D gravity path integral. The higher order terms in the potential are what generate the polygons in the tesselation. From this point of view, the double scaling limit is the process of scaling into the edge to capture the universal physics that survives in taking a continuum limit by  picking out the surfaces that have a diverging number of elements in a tesselation:  at fixed area this is effectively sending the size of the tesselations to zero.

For the 0B cases, the $k$th multicritical model corresponds to having  additional zeros at the origin where the two symmetric components of the distribution of $Q$ eigenvalues ${\tilde y}_i$ collide and merge~\cite{Periwal:1990gf,Periwal:1990qb,Crnkovic:1990mr,
Crnkovic:1992wd,Klebanov:2003km}. The tree-level analysis is similar (in working with $\lambda_i={\tilde y}^2_i$) to the 0A case, but the physics is rather different in general, arising from the fact that the merged two-cut system, once scaled into the region of interest,  treats perturbation theory quite differently from a system resulting from zooming in on an  endpoint. The consequences for 0B ${\cal N}{=}1$ JT supergravity were anticipated in ref.~\cite{Stanford:2019vob} and further explored in ref.~\cite{Johnson:2021owr}, where string equations were used to study features at all orders in perturbation theory as well as non-perturbatively.

\subsubsection{The output of the double scaling limit}
The random matrix models have a useful rewriting in terms of an infinite family of orthogonal polynomials that satisfy a recursion relation. A finite number, $N$,  of these polynomials, can be used as a basis to describe all observables in the matrix model. The potential of the model determines the coefficients in the recursion, and ultimately, string equations are just equations for these recursion coefficients. Knowing those gives the orthogonal polynomials.

In the 0A case, there is one family of polynomials, normalized such that they are of the form $P_n(\lambda){=}\lambda^n{+}\cdots$. With this  choice of normalization, they  satisfy the classic three-term recursion relation of the form: $\lambda P_n{=}P_{n+1}{+}S_nP_n{+}R_nP_{n-1}$. 
For a given matrix model potential $V(\lambda)$, the $S_n$ and $R_n$ (and hence the $P_n(\lambda)$) are fully determined through a family of difference equations that follow from $V(\lambda)$. The  $N\times N$ matrix model's  defining integral itself is built from the first $N$ of the orthogonal polynomials, and any question about the  matrix model can be answered in terms of the $S_n$ and~$R_n$.

Consider how the  aforementioned double scaling limit treats these quantities. 
Now it is the  orthogonal polynomial label $n$ that is traded for 
the coordinate $X={n}/{N}$, (continuous at large $N$, and $0{\leq }X{\leq}1$).   The recursion coefficients $S_n$ and $R_n$ become functions $S(X)$ and $R(X)$. The leading spectral density $\rho_0(\lambda)$ has an integral representation in terms of the leading large $N$ solution for $S(X)$ and~$R(X)$ as follows:
\begin{equation}
\label{eq:integral-representation}
    \rho_0(\lambda) = \frac{1}{\pi}\int_0^1
dX\frac{\Theta\left[4R(X)-(\lambda-S(X))^2\right]}{\sqrt{4R(X)-(\lambda-S(X))^2}}\,.
\end{equation}
This first appeared in ref.~\cite{Dalley:1991jp} (generalizing the classic $S{=}0$ case of ref.~\cite{Bessis:1980ss}), although a recent derivation is given in ref.~\cite{Johnson:2023ofr}, where it is discussed, with examples.  As reviewed there, 
 $\rho_0(\lambda)$ falls to its  smallest value when $S(X)$ and $R(X)$ rise to their largest, which is at $X{=}1$.  Therefore $\rho_0(\lambda)$ has support in the range $\l \in (\l_-,\l_+)$ given by:
\be \label{eqn:eigenvalue_endpoints}
\lambda_\pm=\left.\lr{S(X)\pm2\sqrt{R(X)} }\right\rvert_{X=1}\,.
\ee
The centre of the distribution is at
$S_c\equiv S(X=1)$, with width $4\sqrt{R_c}\equiv 4\sqrt{R(X=1)}$.\footnote{As an illustrative case study, the classic Marchenko-Pastur distribution  $\rho_0(\lambda){=}(\pi\lambda)^{-1}\sqrt{(\lambda_+-\lambda)(\lambda-\lambda_-)}$, where $\lambda_\pm{=}1{+}\frac{\widetilde\Gamma}{2}\pm\sqrt{{\widetilde\Gamma}+1}$, comes from the solutions $S(X){=}X{+}\frac{\widetilde\Gamma}{2}$ and $R(x){=}\frac{X^2}{4}{+}\frac{\widetilde\Gamma}{4}X$.} See figure~\ref{fig:unscaled-density}. 
\begin{figure}[t]
    \centering
    \includegraphics[width=0.75\textwidth]{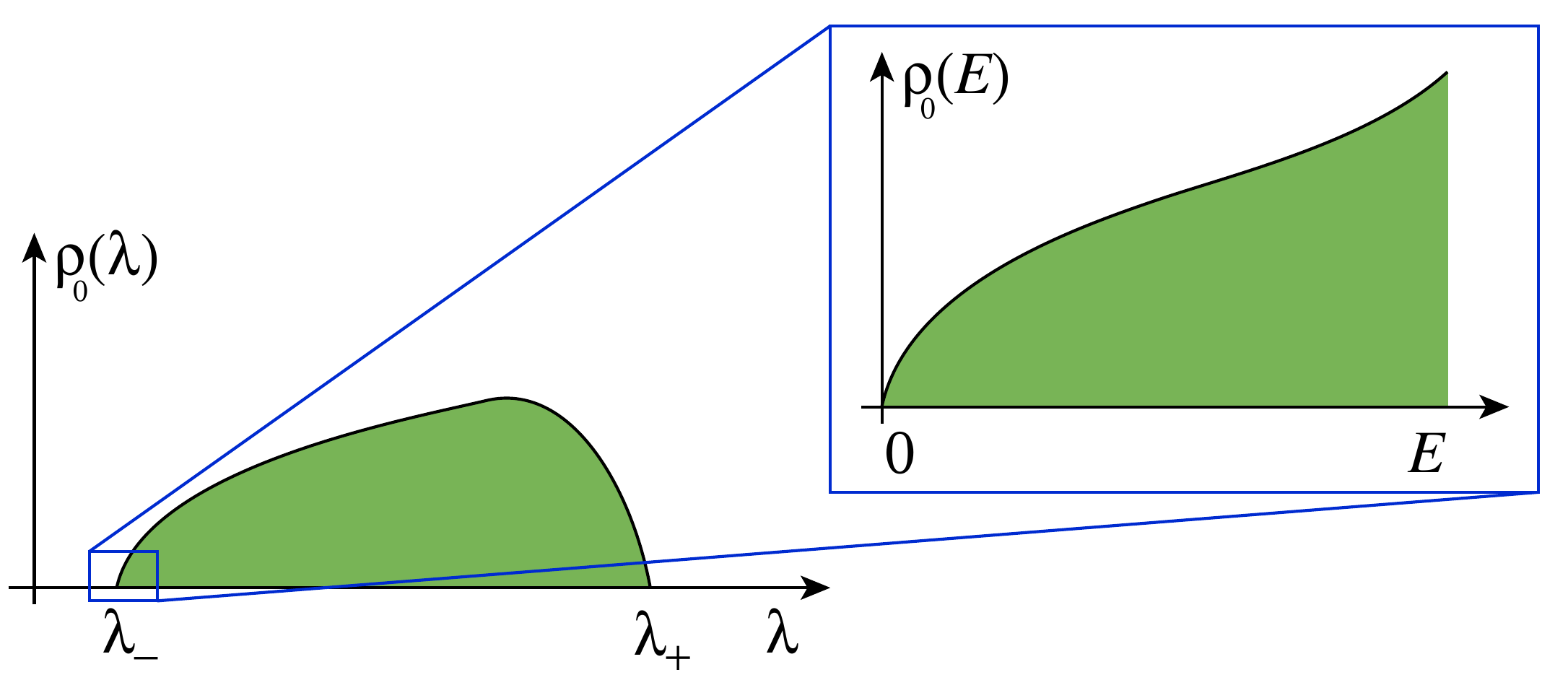}
    \caption{Left: A generic density with  ends at $\lambda_\pm$. Magnifying the infinitessimal region near $\lambda{=}\lambda_-$ recovers universal physics in the double scaling limit.}
    \label{fig:unscaled-density}
\end{figure}
The universal physics to be found in the neighbourhood of one or other endpoint,  can be written in terms of the orthogonal polynomial quantities  that survive the limit.

All that is really needed for the purposes of this paper is that the universal physics of the scaled endpoint is captured by a function $u(x)$,  where $-\infty\leq x\leq+\infty$. 
%
In fact, coordinate $x$ is the infinitessimal  deviation  of $X$ away from $1$, {\it i.e.,} $X=1+(x-\mu)\delta^{2k}$ for some infinitessimal parameter $\delta$ that scales to zero in the double scaling limit as $N{\to}\infty$.\footnote{The parameter $\delta$ can intuitively be thought of keeping track of the typical size of an element of the tesselation dual to the 't Hooft-Feynamn diagrams. The precise powers of $\delta$ that appear in defining the scaling quantities to give non-trivial physics in the $k$th model  ultimately follow from dimensional analysis.} With this parameterization, the region $x<0$ pulls away from the endpoint ($X=1$), and $x{=}\mu$ lands on it.  Also,  $u(x)$ is the scaled part of the combination of $S(X){-}2\sqrt{R(X)}$, which becomes $\lambda_-$ as $X{\to}1$. Carefully approaching the limit by writing $S(X){-}2\sqrt{R(X)} = \lambda_-{+}u(x)\delta^{2}$ in the large $N$ limit of the difference equations yields a non-linear ordinary differential equation (ODE) that determines  $u(x)$, known as a ``string equation'', arising as the limit of the difference equations mentioned earlier for $S_n$ and $R_n$.

The $k$th model comes from a distinct critical potential $V_k(\lambda)$ (coming from tuning to the critical behaviour mentioned before) for which there is particular ODE, but they all have the same form, which is~\cite{Morris:1990bw,Dalley:1992qg,Dalley:1992br}:
\begin{equation}
\label{eq:big-string-equation}
u{\cal R}^2-\frac{\hbar^2}2{\cal R}{\cal R}^{\prime\prime}+\frac{\hbar^2}4({\cal R}^\prime)^2=\hbar^2\Gamma^2\ , \qquad{\rm with }\quad{\cal R}{\equiv}\sum_{k=1}^\infty t_k R_k[u]{+}x\ ,
\end{equation}
where a general form has been written for later reference. A prime denotes an $x$-derivative, and~$\hbar$ denotes the part of the expansion parameter $\frac{1}{N}$ that survives in the scaling limit: $\frac{1}{N}{=}\hbar \delta^{2k+1}$ (in the~$k$th model).  A little more will be said below.

In fact, for the $k$th model, ${\cal R}{\equiv} R_k{+}x$ where $R_k[u]$ is the $k$th Gel'fand-Dikii polynomial (see below) in $u(x)$ and its $x$-derivatives, normalized here so that the purely polynomial part is unity, {\it i.e.,} $R_k{=}u^k+\cdots\# \hbar^{2k-2}u^{(2k-2)}$, where $u^{(m)}$ means the $m$th derivative. The intermediate terms involve mixed orders of derivatives, and every derivative comes with an~$\hbar$. The first few are $R_1[u]{=}u$, $R_2[u]{=}u^2{-}\frac{\hbar^2}{3}u^{\prime\prime}$, and $R_3[u]{=}u^3{-}\frac{\hbar^2}{2}(u^\prime)^2{-} {\tiny \hbar^2}uu^{\prime\prime}{+}\frac{\hbar^4}{10}u^{\prime\prime\prime\prime}$.
General~$R_k[u]$ can be determined by a recursion relation not shown here since it will not be needed. 
These multicritical  models can be used as a basis for describing a wide class of supergravity models which are described quite simply by using (as we shall) the general form~(\ref{eq:big-string-equation}) with ${\cal R}{\equiv}\sum_{k=1}^\infty t_k R_k[u]{+}x$ where the~$t_k$ are coefficients that depend on the model. They control how much of each multicritical model contributes to describing the gravity model in question.\footnote{That  one simply linearly combines the critical models inside ${\cal R}$ to get the string equation might be surprising, but it follows from the manner in which the matrix model potential $V(\lambda)$ determines the recursion coefficients. Additive combinations of potentials enter the difference equations additively. The high degree of non-linearity is encoded in the increased complexity of the $R_k[u]$ themselves.}

Looking ahead, the parameter $\hbar$ is the topological expansion parameter of the model, {\it i.e.,} Euclidean spacetimes of Euler character $\chi{=}2{-}2g{-}b$ (here~$g$ is the number of handles, while~$b$ is the number of boundaries)  have weight $\hbar^{-\chi}$ in the partition sum (equivalently, the gravitational path integral). In the particular JT-like models of gravity we discuss, 
$\hbar\equiv{\rm e}^{-S_0}$ where $S_0$ is the $T=0$ extremal entropy, which multiplies the Einstein-Hilbert term in the gravity action.

As a brief aside, later in the paper a slightly more general possibility will be needed. In fact, there is no need for the ``wall'' (described two paragraphs below~(\ref{eq:partition-functions-A-B})) to be located at $\lambda{=}0$. It can be at some non-zero $\Sigma$  (whose  scaled value is $\sigma$ in the double-scaling limit), in which case the string equation is generalized to~\cite{Dalley:1991yi,Johnson:1992wr}:
\begin{equation}
\label{eq:slightly-bigger-string-equation}
(u-\sigma){\cal R}^2-\frac{\hbar^2}2{\cal R}{\cal R}^{\prime\prime}+\frac{\hbar^2}4({\cal R}^\prime)^2=\hbar^2\Gamma^2\ .
\end{equation}
There are several complementary  intuitive pictures for this. The simplest is that this is the string equation for studying (in the double scaled limit) an ensemble of matrices with lowest eigenvalue given by $\sigma$. Formulating  orthogonal polynomials on the half-line again picks up additional boundary terms~\cite{Dalley:1991xx} when the boundary is at a non-zero position, and the derivation goes through as before. Alternatively, the Dyson gas picture has a logarithmic repulsion term modified to $-\Gamma\sum_i^N \log(\lambda_i-\Sigma)$. A final picture is that one can obtain this by a shift of the backgrounds described by $t_k$, naturally implemented using the ``Virasoro constraints"~\cite{Dalley:1992br,Johnson:1992wr,Johnson:1993vk}.
For now, it will be enough to work with the original case of $\sigma{=}0$, but this more general string equation will be called upon later.

Having solved the full string equation for $u(x)$, and hence obtained the recursion coefficients determining the orthogonal polynomials, the next logical step is that the orthogonal polynomials themselves should be recovered and then used to construct matrix model answers to physics questions in the double scaling limit. This is a beautiful story not needed here except to say that it is all neatly phrased in terms of an auxiliary quantum mechanics system with Hamiltonian ${\cal H}{\equiv}{-}\hbar^2\frac{\partial^2}{\partial x^2}{+}u(x)$. The wavefunctions $\psi(E,x)$ of this Hamiltonian, with energy $E$ play the role of the orthogonal polynomials in the double scaling limit, and $E$ is the value of $\lambda$ away from the endpoint, $\lambda{=}\lambda_-{+}E\delta^2$. This is the process of ``zooming in'' to the endpoint mentioned earlier. See figure~\ref{fig:unscaled-density}.

The JT-like gravity partition function in these models is really the expectation value of a loop of length $\beta$, for which there is a useful represenation in terms of the auxiliary quantum mechanics\cite{Banks:1990df}:
Correlation functions of such operators are naturally computed in the auxiliary quantum mechanics governed by ${\cal H}$. For the expectation value of the most basic such  ``macroscopic loop'' operator the  computation is as follows:
\begin{equation}
\label{eq:basic-loop}
   Z(\beta)\equiv \langle {\rm Tr}(\e^{-\beta H})\rangle=\int_{-\infty}^\mu\! \langle x|\e^{-\beta{\cal H}}|x\rangle\, dx\ .
\end{equation}
This should be read with care: The central part means an average of a particular function of $H$ in the ensemble of matrices $H$, while the right hand side is a computation in the auxiliary quantum mechanics, with Hamiltonian ${\cal H}$. ($H$ and ${\cal H}$ should {\it not}  be confused: The $H$'s have discrete spectra while ${\cal H}$ has a continuous spectrum.  ${\cal H}$ contains {\it statistical} information about all possible $H$ spectra.) The number~$\mu$   will be further discussed below. It is fixed by comparing the leading part of equation~(\ref{eq:basic-loop}) (or its Laplace transform, the leading spectral density) to the leading gravity physics.  This will be done below.

Turning to the 0B case, there is a family of string equations~\cite{Periwal:1990gf,Periwal:1990qb,Crnkovic:1990mr,Crnkovic:1992wd} that can be derived from double-scaling the merging two-cut system as well, describing two families of orthogonal polynomials from which the physics can be recovered. The analogous macroscopic loop can be described in very similar language to that used for the 0A case above. Since we will not use the 0B system much in what is to follow (although at the end of Section~\ref{sec:Neq3-JT-supergravity} we show that one of the ${\cal N}{=}3$ models fits well in this framework), we will not review more of the details of the 0B system henceforth. Ref.~\cite{johnson:2025vyz} contains  recent (relevant) new applications concerning 0B and random matrix models, including new results on the relation between the 0A and 0B  formulations through their respective string equations. There is also a good deal of background presented there, as well as in ref.~\cite{Klebanov:2003km}.

\subsubsection{The leading spectral density}

The full solution  for $u(x)$ admits a perturbative (in $\hbar$) piece and a non-perturbative piece:
 \begin{equation}
     u(x)=u_0(x)+\sum_{g=1}u_g(x) \hbar^{2g}+\cdots\ ,
 \end{equation} 
 where the ellipsis denotes non-perturbative parts. Here, $u_0(x)$ denotes the  (large $N$) leading order  part, obtained by sending $\hbar{\to}0$. The leading spectral density, $\rho_0(E)$, can be derived by taking the scaling limit of the integral representation~(\ref{eq:integral-representation}) which yields:\footnote{The fact that the $x$-integration range is finite corresponds to the aforementioned fact that the  matrix model only uses the first $N$ of the infinite family of orthogonal polynomials.}
\begin{equation}
\label{eq:leading-representation}
    \rho_0(E)=\frac{1}{2\pi\hbar}\int_{-\infty}^\mu \frac{\Theta(E-u_0(x))}{\sqrt{E-u_0(x)}} dx\ .
\end{equation}
Some intuition about how 
this representation 
works is useful here.  There are certain characteristic behaviours of the function $u_0(x)$ that emerge from the matrix model, resulting in certain signature behaviours for $\rho_0(E)$.  Consider first  the case of   $x{<}0$. The function $u_0(x)$ grows to the left (increasing~$|x|$) in this regime (following from the growth of $S(X){-}2\sqrt{R(X)}$ before double scaling).  So in order to get contributions to the integral (given the $\Theta$-function), one has to move to higher energies in order to keep $E{>}u_0(x)$. The resulting energy dependence of $\rho_0(E)$ for increasing $E$ will therefore depend upon the rate at which $u_0(x)$ grows with increasingly negative~$x$. Looking at the leading piece of the  string equation~(\ref{eq:big-string-equation}) (setting $\hbar{=}0$), the  $k$th multicritical universality class mentioned earlier has  $u_0(x)$ solve the leading equation $u_0^k{+}x{=}0$ for $x{<}0$ (up ahead, equations~(\ref{eq:leading-Neq-1-string-equation}) and~(\ref{eq:ansatz-1}) make this explicit) and so  $u_0{\sim} (-x)^{1/k}$.   The integral~(\ref{eq:leading-representation}) shows that this results in the growth $\rho_0{\sim} E^{k-\frac12}$. This is precisely the Wigner square root  joined by $k-1$ extra zeros  for the edge fall-off, as described in Section~\ref{sec:multicriticality}. Ultimately, the kinds of spectral density arising from  a general JT gravity can be interpreted as coming from a mixture of behaviours corresponding to different values of $k$, so the leading  $x{<}0$ equation is ${\cal R}_0\equiv \sum_{k} t_k u_0^k+x=0$, for some numbers~$t_k$, resulting,  through (\ref{eq:leading-representation}) in a decomposition of the form $2\pi\hbar \rho_0(E)=\sum_{k=1}C_kE^{k-\frac12}$.  The~$C_k$ are numbers that follow from the~$t_k$.

Recall that $x{=}\mu$ corresponds to $X{=}1$, where the classical edge of the density $\rho_0(E)$ is located. The value of the energy there is $E_0\equiv u_0(\mu)$. In fact, it is very useful to change variables from $x$ to $u_0$ in the integral representation~(\ref{eq:leading-representation}), giving:
\begin{equation}
\label{eq:leading-representation-u0}
    \rho_0(E)=\frac{1}{2\pi\hbar}\int_{E_0}^E \frac{f(u_0)}{\sqrt{E-u_0}} du_0\ ,
\end{equation}
where $f(u_0){=}{-}\partial x/\partial u_0$ is (minus) the Jacobian for the change of variables. For ordinary multicritical  models (of application to ordinary JT gravity), the value of $\mu$ is typically zero and the string equation for the entire $x<0$ regime is $\sum t_k u_0^k+x=0$ and hence $E_0{=}u_0(0){=}0$.  Non-zero $E_0$ will appear shortly, however, and will be typical in the models that are the subject of this paper.

\subsection{JT supergravity theories as  multicritical matrix models}
\label{sec:JT-supergravity-multicritical}
Recall that for the study of supergravity, the  random matrix models are ensembles of positive matrices, and so the spectrum stops at $E{=}0$. Therefore the function $u_0(x)$ will stay positive on the real $x$ line, so that $E<0$ can never contribute (given the presence of the $\Theta$-function in the integral~(\ref{eq:leading-representation})). Since $u_0(x)$ is decreasing as one moves in from the left, it natural for it to asymptote to zero (or more generally some positive value) as one moves to the far right in $x$.\footnote{This is quite different from more familiar random matrix models such as the GUE prototype where a non-perturbative tail for $E<0$ is present. There, $u_0(x)$ can naturally  go negative with increasing positive $x$, and there are viable $\psi(E,x)$ wavefunctions for negative $E$ whose effects appear semi-classically and fully non-perturbatively as the tail. } Rising to the asymptotic far right is problematic since it would result from a system with  a leading energy spectrum that is discrete, which is not the case here.\footnote{Non-perturbatively, things are more subtle. The full $u(x)$ can contain regions where it falls and rises again to the right (asymptoting to zero), but the work of ref.~\cite{Carlisle:2005mk}  strongly suggests that the resulting wells do not support any discrete  ({\it i.e.} bound) states.}
\subsubsection{${\cal N}{=}1$ JT supergravity}
\label{sec:N_eq_1_supergravity}
The simplest  natural solution in the  positive $x>0$ regime is simply $u_0(x)=0$, and indeed this readily emerged in studies of ${\cal N}{=}1$ JT supergravity~\cite{Johnson:2020heh,Johnson:2020exp}. The leading string equation is simply: 
\begin{equation}
\label{eq:leading-Neq-1-string-equation}
    u_0{\cal R}_0^2=0 \ ,\qquad{\rm with}\quad {\cal R}_0[u_0]\equiv\sum_{k=1}^\infty t_k u_0^k+x\ ,
\end{equation} and a summary of the leading solution is:
\begin{eqnarray}
\label{eq:ansatz-1}
{\cal R}_0[u_0] &=& 0\ ,\qquad x\leq0\ ,\\
    u_0(x)&=&0\ ,\qquad x>0 \ ,
\end{eqnarray}
so the complete leading solution for $u_0$ is constructed piecewise. 
For these models, the upper value of the integration, $\mu$, is positive (the precise value depends on conventions), and allied to that is the fact that the  key role of the $u_0=0$ portion of the solution is to contribute, {\it via} equation~(\ref{eq:leading-representation}), the tail $\rho_0\sim\frac{\mu}{2\pi\hbar}\frac{1}{E^{1/2}}$, characteristic of Wishart-type models. See  figure~\ref{fig:super-density}(a).
\begin{figure}[t]
    \centering
    \includegraphics[width=0.85\textwidth]{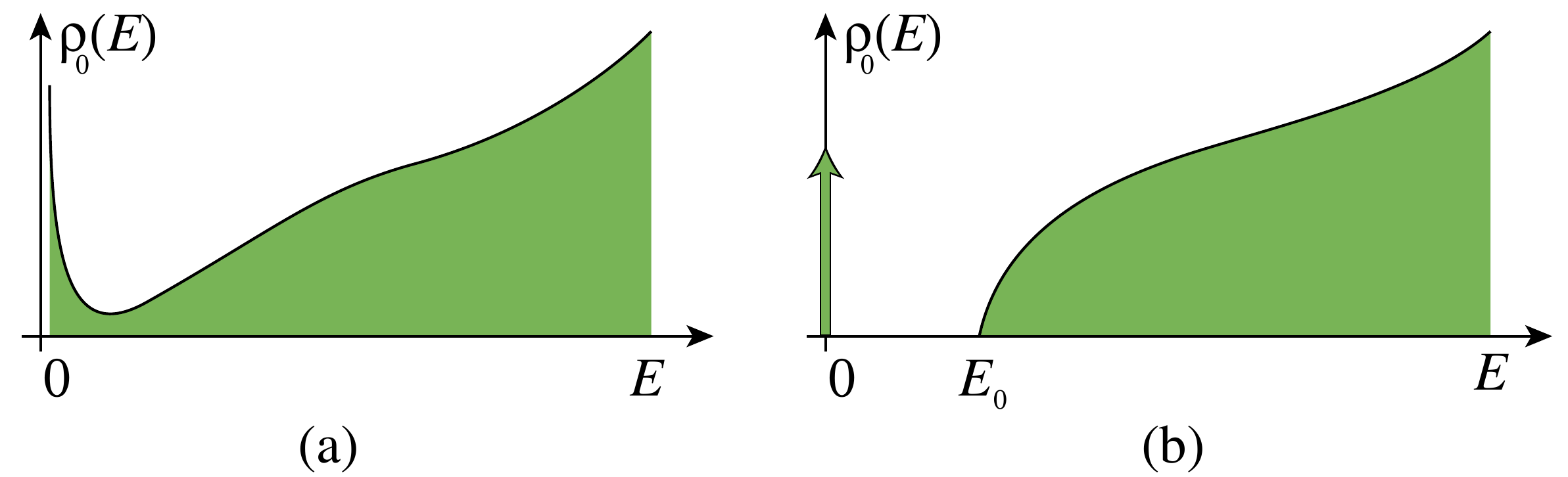}
    \caption{(a) The typical ${\cal N}{=}1$ behaviour of a density. (b) The possible ${\cal N}{>}1$ situation with some BPS states at $E{=}0$ and a non-BPS sector beginning at some threshold energy $E_0$.}
    \label{fig:super-density}
\end{figure}

All the elements are in place to fix the form of the leading string equation's solution, using the spectral density from ${\cal N}{=}1$ JT supergravity~\cite{Stanford:2017thb,Stanford:2019vob} as input. The idea~\cite{Johnson:2020heh,Johnson:2020exp} is simply to expand $\rho_0(E){=}\cosh(2\pi\sqrt{E})/2\pi\hbar\sqrt{E}$ in powers of~$E$ as $\rho_0(E){=}C_0{E^{-\frac12}}+\sum_{k=1} C_k E^{k-\frac12}$, and identifying the $k$th term as coming from some admixture of the $k$th multicritical model. This uniquely fixes the $t_k$ in the string equation to be $t_k {=} \pi^{2k}/(k!)^2$. The fact that $C_0{=}1/2\pi\hbar$ fixes $\mu{=}1$, and with these $t_k$, the resulting leading string equation in the $x{<}0$ regime resums to give $I_0(2\pi\sqrt{u_0}){-}1{+}x{=}0$. 

\subsubsection{Extended JT supergravity and a Multicritical Miracle}
\label{sec:the-miracle}
Turning to models of extended JT supergravity, the focus of this paper,  it is natural to ask\cite{Johnson:2023ofr} if they can also be built as a combination of multicritical models for some $t_k$. A particularly striking feature that emerges is the following. Now the spectrum~\cite{Stanford:2017thb,Mertens:2017mtv,Turiaci:2023jfa} can come in two distinct kinds of sectors, non-BPS states which form a continuum starting at some finite non-zero energy (say $E_0$), and BPS states which are $\delta$-function supported at a particular energy, for example $E{=}0$. See figure~\ref{fig:super-density}(b). 

The observation of ref.~\cite{Johnson:2023ofr} was that there is a refinement of the above behaviours that can naturally describe all of this in this class of matrix models. It can be  arrived at as follows: 
While generically $u_0(x){=}0$ in the $x>0$ region, at the next order in $\hbar$ perturbation theory, the leading correction to $u_0(x)$ is of the form $\hbar^2\Gamma^2/x^2$, and signals the  presence of some number, $\Gamma$, of degenerate $E=0$ states of the aforementioned Schr\"odinger problem for which $u(x)$ is a potential~\cite{Carlisle:2005wa,Carlisle:2005mk}. If instead $\Gamma$ scales as~$\hbar^{-1}$,  this term is leading in  small $\hbar$ perturbation theory (this is the double-scaled version of the scaling discussed below equation~(\ref{eq:dyson-gas-0A})). Put differently, this would mean that there are $\Gamma{\sim}{\rm e}^{S_0}$ zero energy states in the problem, precisely what is needed for the kinds of gravity problem in hand. 

The ``backreaction'' from such a large number of states  must be taken into account in the leading solution for $u_0(x)$. This suggests that instead of the  exact $u_0(x){=}0$ solution for $x{>}0$ as in~(\ref{eq:leading-Neq-1-string-equation}), the ansatz~\cite{Johnson:2023ofr}:
\begin{eqnarray}
{\cal R}_0[u_0] &=& 0\ ,\qquad x\to-\infty \nonumber\\
    u_0(x)&=&\frac{\widetilde\Gamma^2}{x^2}\ ,\qquad x\to+\infty\ ,
\label{eq:the-ansatz}
\end{eqnarray}
should be used instead, where ${\widetilde\Gamma}{=}\hbar\Gamma$ is held finite as $\hbar\to0$. (These are to be understood as approximate solutions for large $|x|$, which is all that is needed to capture the physics in a large $E$ expansion. Section~\ref{sec:new-approach} will discuss this more.) This behaviour can all be nicely accommodated in a ${\widetilde\Gamma}$-deformation of the previous leading equation~(\ref{eq:leading-Neq-1-string-equation}):
\begin{equation}
\label{eq:leading-Neq-2-string-equation}
    u_0{\cal R}_0^2={\widetilde\Gamma}^2 \ ,
\end{equation}
which is indeed what comes from  the string equation~(\ref{eq:big-string-equation}) for this class of models. In this way, one still has the elements needed for a continuum piece of the spectral density ($u_0(x)$ rising to the left and falling to the right), a 
 supersymmetric model ($u_0(x)$ positive), but now incorporating some 
 $\Gamma{\sim}{\rm e}^{S_0}$ 
 zero energy states.
Note that the positive $x$ integral for $u_0(x)$ with this ansatz yields more general behaviour. Notably, since at $x{=}\mu$ the value of $u_0$ must  be the threshold energy  $E_0$, we also have $u_0(x){=}\mu^2 E_0/x^2$, and hence $\widetilde\Gamma {=}\mu \sqrt{E_0}.$
The connection (through the ansatz) between $\Gamma$ and the value of the threshold energy is the first hint that the string equation links the BPS and non-BPS sectors. In fact, a much stronger and striking  connection  emerges.


This  is where something seemingly miraculous occurs.  While the ansatz~(\ref{eq:the-ansatz}) can now accommodate the disc order contributions of both BPS and non--BPS sectors, there are several reasons why it could simply fail to work, and spectacularly so. Starting with the particular form of the non-BPS continuum, ({\it e.g.} $\rho_0(E){=}\e^{S_0}\sin(2\pi\sqrt{E-E_0})/(8\pi^3E)$ for ${\cal N}{=}2$) ref.~\cite{Johnson:2023ofr}  developed a formal expansion of it in small $E_0/E$. This  yields an infinite set of both positive and {\it negative} powers of $E^\frac12$. The coefficients of the negative powers have two sources. One is the presence of the threshold energy: The $k$th constituent multicritical model coming from the behaviour of $u_0(x)$ in the $x<0$ regime can be expected to produce  terms involving products with  $(E-E_0)^\frac12$, (a precise formula is reviewed below) which yields such negative powers upon expanding. The other source is the form of the special ansatz for $u_0(x)$ in the  $x>0$ regime, which after integrating and   expanding, also produces such terms.  Given the positive powers, with some care, a set of  $t_k(E_0)$ can be found that accounts for them, and correspondingly this allows for some of the contributions to the infinite series of negative powers of $E^\frac12$ to be attributed to individual multicritical models. But  once one also isolates the contributions coming from the $x>0$ ansatz~(\ref{eq:the-ansatz}),  {\it all other contributions} to the infinite family of negative powers must vanish. 

This places a constraint on the only adjustable parameter left, which is  $\mu$ (equivalently, $\Gamma$, through the reasoning below~(\ref{eq:leading-Neq-2-string-equation})), which must be some particular function of $E_0$ for this to work. 
The remarkable outcome is that this infinite family of unmatched coefficients indeed vanishes if and only if the dependence of $\Gamma$ on the parameters of the theory obey the BPS formula ({\it e.g.,} $\Gamma{=}{\rm e}^{S_0}\sin(2\pi\sqrt{E_0})/4\pi^2$ for ${\cal N}{=}2$.)  A similar success was reported for this scheme as applied to the (small) ${\cal N}{=}4$ case in ref.~\cite{Johnson:2024tgg}. In other words, requiring the multicritical matrix model to reproduce the non-BPS sector {\it only works} if it also has  precisely the correct BPS sector! 

From some perspectives this is very surprising, since one might have not suspected that the matrix model has any deep knowledge about the inner workings of supersymmetry to allow it to forbid the inclusion of solutions that don't combine BPS and non-BPS in the correct way. In retrospect, one might argue that this is natural: On the gravity side, the required  supersymmetry type is implemented by summing over surfaces with the correct spin structures, hence including contributions from the appropriate kinds of supermanifold. From the 't Hooftian perspective, a random matrix model is also  a consistent summation over some class of 2D manifolds (although how the correct supermanifold contributions  are extracted by the double scaling limit is both mysterious and miraculous, but it does it). Moreover, once one fixes the leading  ({\it i.e.,} disc order) physics, everything else follows from recursive properties of the matrix model to higher order in the genus expansion (due to the loop equations,  topological recursion, and so forth). So  already at  disc level, matrix model consistency should  correctly tie together all the elements that follow  from the implicit sum over underlying random surfaces. 

That having been said, it is all still a rather miraculous outcome, and together these results for ${\cal N}{=}2$ and  (small) ${\cal N}{=}4$ (as well as the connections between gaps and eigenvalue repulsion~\cite{Johnson:2024tgg}) can be taken as evidence for the robustness of the correspondence between the gravity path integral for JT gravity and random matrix models that go well beyond the bosonic successes. 

 As already stated in the opening remarks, the goal of this paper is to get better understanding of the miracle just reviewed, by refining the methodology for using the ansatz~(\ref{eq:the-ansatz}), through the string equation,  to seek multicritical matrix model descriptions of a wider range of extended JT supergravity spectra. A key aspect of this will be to replace the (implicit) large~$E$ expansion described above by an expansion that allows exploration of the leading behaviour all the way down to~$E{=}0$. Once this has been done, the technique can be applied to more examples, and perhaps used to anticipate and constrain others to come.

\section{A New Approach}
\label{sec:new-approach}

It is worth first addressing some of the details of the ``miracle'' approach reviewed in the last section, in order to identify what aspects need a stronger light shone on them. There were two key  ingredients. 
One is from the $x{>}0$ asymptotic of the ansatz~(\ref{eq:the-ansatz}) where $u_0(x)={\widetilde\Gamma}^2/x^2 = {\mu^2 E_0}/x^2$, where the latter equality comes from the condition of setting the value of the threshold energy to be $u_0(\mu)$.  Putting this into the integral (either form~(\ref{eq:leading-representation}) or form~(\ref{eq:leading-representation-u0})) yielded  $\frac{\mu}{2\pi}\sqrt{E-E_0}/E$.  Expanding this  produces   negative  powers of $E^\frac12$.
The other ingredient  was that following from the $x<0$ part of the ansatz~(\ref{eq:the-ansatz}): the spectrum should be composed of  spectra produced by individual multicritical models. This would have come from putting Jacobian $f(u_0){=}t_k k u_0^{k-1}$ into integral~(\ref{eq:leading-representation-u0}), yielding the general result derived in ref.~\cite{Johnson:2023ofr}:
\begin{equation}
\rho_0^{(k)}(E_0,E)={\widetilde C}_k(E_0,E)E^{k-\frac12}/2\pi\hbar\ , 
\end{equation}
where, defining $\cos\theta_0{=}\sqrt{1-{E_0}/{E}}$, 
\begin{eqnarray}
\label{eq:cee_kay_big}
{\widetilde C}_k(E_0,E)= \frac{2k}{2^{2k-2}}\sum_{i=1}^k\frac{(2k-1)!(-1)^{i-1}}{(k-i)!(k+i-1)!}\frac{\cos\left[(2i-1)\theta_0\right]}{(2i-1)}\ .
\end{eqnarray}
For clarity, the first few cases  for  $k$  have: 
\begin{eqnarray}
\label{eq:examples-of-minimals}
\rho_0^{(1)}\! &=& \!2\frac{(E-E_0)^\frac12}{2\pi\hbar}\ ,\,\, \rho_0^{(2)} = \frac83\frac{(E-E_0)^\frac12}{2\pi\hbar}\left(E+\frac12 E_0\right)\ ,
\nonumber\\
\rho_0^{(3)}\! &=&\! \frac{16}{5}\frac{(E-E_0)^\frac12}{2\pi\hbar}\left(E^2+\frac12EE_0+\frac38E_0^2\right)\ .
\end{eqnarray}
It is in expanding in small $E_0/E$ that these $\rho_0^{(k)}(E_0,E)$ also generate an infinite number of negative powers of $E^\frac12$.
Notice that the threshold energy $E_0$ must be common to each sector ({\it i.e.,} $x{<}0$ and~$x{>}0$), and it is equal to $u_0(\mu)$. That value of $\mu$ must lie in the positive $x$ sector for both cases. To get the usual equation for the multicritical models to yield such a threshold, one can easily modify  the  $u_0(x)$ equation with a constant shift  to give:
$ \sum t_k u_0^k+x-\mu- \sum t_k E_0^k=0$. Notice that this  extends the curve into the  $x{>}0$ domain beyond $x{=}\mu$. This is precisely what gives the desired spectral density, as it yields the correct Jacobian and the $E_0$ lower limit for integral (\ref{eq:leading-representation-u0}).

But  of course  this extension into the $x{>}0$ domain  contradicts the fact that the $x{>}0$ inverse square form should have already taken over already in that region before $x=\mu$. The key point here is that these two pieces of input for $u_0(x)$ in ansatz~(\ref{eq:the-ansatz}) {\it cannot} be the correct result for all energies, but they work well enough for the validity of the expansion being used, which is for large enough $E$ such that $E_0/E$ is a good expansion parameter. For low enough $E$ (where $|x|$ is no longer large) these forms of $u_0(x)$ clash with each other, which of course also fits with the fact that in the above discussion  we modified the ${\cal R}_0$ equation by hand.

The point here is that there is no real contradiction since at low enough energies the solution for $u_0(x)$ in the intermediate~$x$ region is of neither functional form given in the ansatz~(\ref{eq:the-ansatz}). Those useful forms worked well precisely because the $E_0/E$ expansion allowed their shortcomings in the interior to be ignored. The miracle result connecting all their contributions together to give a consistent multicritical expansion (on condition that the BPS and non-BPS sectors are precisely connected) was controlled only by the asymptotic parts of these forms of $u_0(x)$.\footnote{In  retrospect this makes the miracle perhaps even more miraculous.}

This makes it clear that  full  smooth solution $u_0(x)$ of the leading string equation~(\ref{eq:leading-Neq-1-string-equation}) has yet to be explored, and that is what clearly should be done next. Writing it out more explicitly:
\begin{equation}
    u_0 \left(\sum_{k=1} t_k u_0^k+x \right)^2 = \tilde{\Gamma}^2\ , \label{eq:stringeq}
\end{equation}
with ${\widetilde\Gamma}=\hbar\Gamma$, where $\hbar\equiv{\rm e}^{-{S_0}}$. Solving this  equation for $x$ worked well in the past, but in the presence of $\widetilde\Gamma$ that is no longer convenient. However it is not needed to make progress, since all we need is $f(u_0) {=} {-}dx/d u_0$ for use as a Jacobian in the second form of the integral representation~(\ref{eq:leading-representation-u0}).  To this end,  differentiate \eqref{eq:stringeq} with respect to $u_0$, yielding:
\begin{equation}
    0 = \left(\sum_{k=1} t_k u_0^k+x \right)^2 + 2u_0 \left(\sum_{k=1} t_k u_0^k+x \right) \left(\sum_{k=1} t_k k u_0^{k-1}-f(u_0) \right).
\end{equation}
We then use~(\ref{eq:stringeq}) to  rewrite $\left(\sum_{k=1} t_k u_0^k+x \right)$ in terms of $\widetilde\Gamma$ to obtain the following simple expression for $f(u_0)$:
\begin{subequations}
\begin{equation}
\label{eq:Jacob-result}
    f(u_0) = f_0 \pm \frac{|\widetilde\Gamma|}{2 u_0^{3/2}}\ ,
\quad 
{\rm where} 
\quad
    f_0 = \sum_{k=1}t_k k u_0^{k-1}\ .
\end{equation}
\end{subequations}
It is important to note that we have two different choices for $f(u_0)$, since the string equation is quadratic in $x$. Insert $f(u_0)$ back to the integral representation~\eqref{eq:leading-representation-u0} yields a nice form:
\begin{equation}
\label{eq:nice-form}
    \rho_0(E) = \frac{1}{2\pi \hbar} \int_{E_0}^E \frac{f_0(u_0) du_0}{\sqrt{E-u_0}}  \pm \frac{|\widetilde\Gamma|}{4\pi \hbar} \int_{E_0}^E \frac{du_0}{u_0^{3/2} \sqrt{E-u_0}}\ , 
\end{equation}
which are integrals that we have already encountered in the discussion so far. The point here is that working explicitly with~$u_0$ as the integration variable gives  a form for the density that  can be trusted to have  come from the two kinds of ingredient in the ansatz~(\ref{eq:the-ansatz}), while at the same time ensuring that they must smoothly and consistently hand over to each other (as $x$ is traversed) in the way governed by the leading string equation~(\ref{eq:leading-Neq-2-string-equation}). 

Having taken care that the leading string equation is now reliably incorporated for all $x$,  it is time to make a more powerful choice of expansion variable. Replace $E$ according to: 
\begin{equation}
    E = E_0 (1+ \lambda)\ ,
\end{equation}
and in terms of new variables\footnote{Here, the symbol $\lambda$ is not intended to imply a connection to the unscaled eigenvalues discussed in Section~\ref{sec:string-equations-general}.} $(\lambda, E_0)$, the first integral in (\ref{eq:nice-form}) is made of summands of the form:
\begin{subequations}
    \begin{equation}
        \int_{E_0}^E \frac{u_0^{k-1} du_0}{\sqrt{E-u_0}} = 2 E_0^{k-1/2} \sqrt{\lambda}\  {}_2 F_{1} \left(1-k, 1, \frac{3}{2}, -\lambda\right)\ .
    \end{equation}
    Since, when not zero, its first argument is a non-positive integer, the  function $_2F_{1} \left(1-k, 1, \frac{3}{2}, -\lambda\right)$ yields  a polynomial of degree $k-1$ in $\lambda$, which will be denoted as $w_k (\lambda)/2$. It is related to ref. \cite{Johnson:2023ofr}'s~${\widetilde C}_k$ polynomials~(\ref{eq:cee_kay_big}) as follows:
    \begin{equation}
        w_k(\lambda) = \frac{(1+\lambda)^{k-1/2}}{k \sqrt{\lambda}} \widetilde{C}_k (1, 1+\lambda)\ .
    \end{equation}
    For future reference, let us notice that it can be written as:
    \begin{eqnarray}
    \label{eq:w_k-lambda}
      w_k(\lambda) = 2  \sum _{n=0}^{k-1} \frac{(4 \lambda )^n (k-1)! n!}{(2 n+1)! (k-n-1)!} \ .
    \end{eqnarray}
    From the second integral in~(\ref{eq:nice-form}):
    \begin{equation}
 \int_{E_0}^E \frac{du_0}{u_0^{3/2} \sqrt{E-u_0}}=\frac{2 \sqrt{\lambda}}{E_0 (1+\lambda)} = \frac{2\sqrt{E-E_0}}{\sqrt{E_0}}\frac{1}{E}\ .
    \end{equation}
\end{subequations}
While the first integral was regular everywhere,  this second one has a simple pole at $\lambda {=} {-}1$, {\it i.e.,} at $E{=}0$. There is also  a common branch cut along $E{-}E_0>0$.\footnote{We presume that this analytic structure is not changed upon performing the sum in $f= \sum_k t^k k u_0^{k-1}$. Note that this puts a tight constraint on the admissible forms of $\rho_0(E)$.} Finally then, equation~(\ref{eq:nice-form}) can be rewritten as:
\begin{equation}
\label{eq:representation-new}
    \rho_0(E) = \frac{1}{2\pi \hbar} \sum_{k=1}t_k k E_0^{k-1/2} \sqrt{\lambda} w_k (\lambda)   \pm \frac{|\widetilde\Gamma|}{2\pi \hbar} \frac{ \sqrt{\lambda}}{E_0 (1+\lambda)}\ . 
\end{equation}
The general plan is now in two simple steps:
\begin{enumerate}[]
\item {\bf Step 1:} When comparing to a specific $\rho_0(E)$ ({\it i.e.} $\rho_0(\lambda)$),  expand both sides in \eqref{eq:representation-new} around $\lambda {=} {-}1$, and pay attention to the behaviour in the complex plane.  Reading off the coefficient of the simple pole yields the exact form of the function $\widetilde\Gamma(E_0)$! Explicitly:
\begin{equation}
  \pm |\widetilde\Gamma(E_0)| = - \hbar E_0 \oint_{\lambda=-1}\rho_0 (\lambda, E_0)\ .\label{eq:Cauchy}
\end{equation}
The choice of sign is determined by the requirement that $|\widetilde\Gamma|$ remains non-negative.\footnote{The reader better-versed with the spectral curve formalism may note that an analogous expression appeared  in ref.~\cite{Turiaci:2023jfa} (see equation~(B.8) therein), arising from the addition of logarithmic terms to the matrix potential. We thank Joaquin Turiaci for pointing this out to us.}  We shall see later that the properties of the matrix model depend crucially on this sign.

\item {\bf Step 2:} Put the result for $\widetilde\Gamma$ back into \eqref{eq:representation-new} and then expand everything around $E_0 {=} 0$. In particular, write:
\begin{equation}
\label{eq:teekay-sum}
    t_k(E_0) = \sum_{n=0} t_{k,n}E_0^n\ ,
\end{equation}
and move the $\widetilde\Gamma$ terms to the left. In all the examples considered below, expanding then in powers of~$E_0$  give  Taylor coefficients of the form
 $   \sqrt{\lambda} W_n (\lambda)$,
where $W$ is a polynomial of degree $n$. $W$'s coefficients are given by $\rho_0(\lambda)$ and $\widetilde\Gamma(E_0)$. Demanding that polynomials on both sides are equal up to a constant uniquely determines $t_{k,n-k}$. Resumming~(\ref{eq:teekay-sum}) yields the needed $t_k(E_0)$.

\end{enumerate}

This completes the process of determining the particular multicritical mixture needed to capture a particular spectral density as a matrix model {\it and} determine the details of the BPS sector. This refined procedure makes it very clear that since it all follows from the single function~(\ref{eq:representation-new}), the BPS and non-BPS sectors are inextricably linked. That form in turn came from the specific form of $u_0$ dictated by the string equation of the underlying matrix model.

\section{${\cal N}{=}2$ JT Supergravity}
\label{sec:Neq2-JT-supergravity}
Let us demonstrate the effectiveness of the refined procedure on concrete examples, starting with $\mathcal{N}{=}2$ JT supergravity~\cite{Fu:2016vas,Stanford:2017thb,Mertens:2017mtv}. The key point~\cite{Turiaci:2023jfa} is that the various contributions to the partition function can be organized into multiplets that are taken to be statistically independent and as such each one should admit its own matrix model description. Each non-BPS multiplet is built out of states with R-charges $q \pm \hat{q}/2$, where $\hat q$ is the R-charge of the supercharge.  BPS multiplets have  charges $k = q\pm\frac{\hat{q}}{2}$, where $\pm$ is a minus sign of $q$ and $|q|\leq {\hat q}$.  The  non-BPS  density of states reads:
\begin{equation}
    \rho_0(E, E_0) = \frac{1}{2\pi \hbar}
\frac{\sinh\left(
2\pi \sqrt{E-E_0}
\right)}{4\pi^2 E} \Theta \left(
E-E_0
\right)\ , \quad {\rm where} \quad E_0 = \frac{q^2}{4\hat{q}^2}\ .
\end{equation}
 On the other hand, BPS multiplets (with R-charge $q {\pm} \frac{\hat{q}}{2}$) have the (disc) density of states:
\begin{equation}
    \rho_0^{\rm BPS} = \frac{1}{4\pi^2 \hbar} \sin \left(
    \frac{\pi q}{\hat{q}}
    \right) \delta(E) \Theta \left( \hat{q}-|q| \right)\ .
\end{equation}
Analytically extending $\rho_0(E, E_0)$ to the whole real line and rewriting it in terms of $(\lambda, E_0)$ gives:
\begin{equation}
    \rho_0(\lambda, E_0) = \frac{1}{2\pi \hbar} \frac{\sinh \left(
    2\pi \sqrt{E_0 \lambda}
    \right)}{4\pi^2 E_0 (1+\lambda)}\ .
\end{equation}
Using \eqref{eq:Cauchy} immediately yields:
\begin{equation}
    \widetilde\Gamma(E_0) = - \hbar E_0 \times \frac{i}{\hbar} \frac{\sinh \left(
    2\pi \sqrt{-E_0}
    \right)}{4\pi^2 E_0} = \frac{\sin \left(2 \pi \sqrt{E_0}\right)}{4\pi^2} = \frac{\sin \left(  \frac{\pi q}{\hat{q}} \right)}{4\pi^2}\ .
\end{equation}
Note that this is exactly the answer for the BPS sector. It was not input into the process, but rather simply follows from the non-BPS density of states! It bears repeating that this is guaranteed by the structure of the underlying string equation which resulted in the form~(\ref{eq:representation-new}).

Following Step 2, it is now convenient to move the term involving $\widetilde\Gamma$ to the other side. Thus, on the left we have
\begin{equation}
    {\rm LHS} = \frac{1}{2\pi \hbar}\frac{1}{4\pi^2 E_0 (1+\lambda)} \left(
\sinh\left(
2\pi \sqrt{E_0 \lambda}
\right) - \sqrt{\lambda} \sin \left(2 \pi \sqrt{E_0} \right)
    \right)\ .
\end{equation}
Expanding the expression in the bracket, one obtains:
\begin{equation}
    {\rm LHS} = \frac{\sqrt{\lambda}}{2\pi \hbar}  \sum_{n=1} E_0^{n-1/2} \frac{(2\pi)^{2n-1}}{(2n+1)!}(-1)^{n-1} \sum_{k=0}^{n-1}(-\lambda)^k\ .
\end{equation}
 On the other hand, on the right:
\begin{equation}
    {\rm RHS} = \frac{\sqrt{\lambda}}{2\pi \hbar} \sum_{n=1}  E_0^{n-1/2} \sum_{k=0}^{n-1} (n-k) t_{n-k,k} w_{n-k}(\lambda)\ .
\end{equation}
Comparing the coefficients in front of various powers of $E_0$ gives:
\begin{equation}
    \frac{(2\pi)^{2n-1}}{(2n+1)!}(-1)^{n-1} \sum_{k=0}^{n-1}(-\lambda)^k = \sum_{k=0}^{n-1} (n-k) t_{n-k,k} w_{n-k}(\lambda)\ .
\end{equation}
Both sides are polynomials in $\lambda$ of order $n-1$. Thus, they will be equal if all powers of $\lambda$ are matched. We may solve it systematically starting from $\lambda^{n-1}$ (this determines $t_{n,0}$) and then knowing $t_{n,0}$ we may obtain $t_{n-1,1}$ by looking at the coefficient in front of $\lambda^{n-2}$ and so on. This procedure yields:\footnote{A  general formula can be derived for the $t_k$ as a series of $\lambda$-derivatives of $\rho$, which may be useful for future applications. It is given in Appendix~\ref{app:teekay-machinery}.}
\begin{equation}
    t_{k,n}= \frac{\pi^{2(k+n)-1} (-1)^n}{2k! (2k+1) n! (n+k)!}\ ,
\end{equation}
which can be resumed to give:
\begin{equation}
\label{eq:teekay_Neq2}
    t_k = \frac{\pi ^{k-1} E_0^{-\frac{k}{2}} J_k\left(2 \pi\sqrt{E_0}  \right)}{2k!(2k+1)}\ ,
\end{equation}
where $J_k$ is the $k$th Bessel function of the first kind.
This answer exactly agrees with the  one derived in ref.~\cite{Johnson:2023ofr}.

Note that we were a little cavalier in going from equation~\eqref{eq:leading-representation} to  equation~\eqref{eq:leading-representation-u0}. It was implicitly assumed that the function $x(u_0)$ is monotonic, {\it i.e.}, that $f(u_0) {=} {-} \frac{dx}{du_0}$ remains positive on the image of~$u_0$ (which lies on $(0, \infty)$), as happens in figure~\ref{fig:u-plots-1}(a). A direct computation shows that this is not always the case, however. 
The  borderline case is $E_0{=}\frac{1}{4}$, where $\widetilde\Gamma$ first vanishes, {\it precisely at the edge of  the range where there are BPS states}. See  figure~\ref{fig:u-plots-1}(b).
When $1>E_0>\frac{1}{4}$, we are led to a minus sign in front of $|\widetilde{\Gamma}|$. As a result, $f(u_0)$ changes sign on $(0, \infty)$. See figure~\ref{fig:u-plots-2}(a).  However, it remains positive on $[E_0, \infty)$, thus keeping equation \eqref{eq:leading-representation-u0} valid.  If $4>E_0>1$, we choose again the plus sign. However,~$u(x)$ is still multivalued (see figure~\ref{fig:u-plots-2}(b)). Thus, also in this case $f$ changes sign (but does it outside the interval $[E_0, \infty)$). For larger $E_0$, either of these two scenarios can take place.
\begin{figure}[t]
    \centering
    \includegraphics[width=0.49\textwidth]{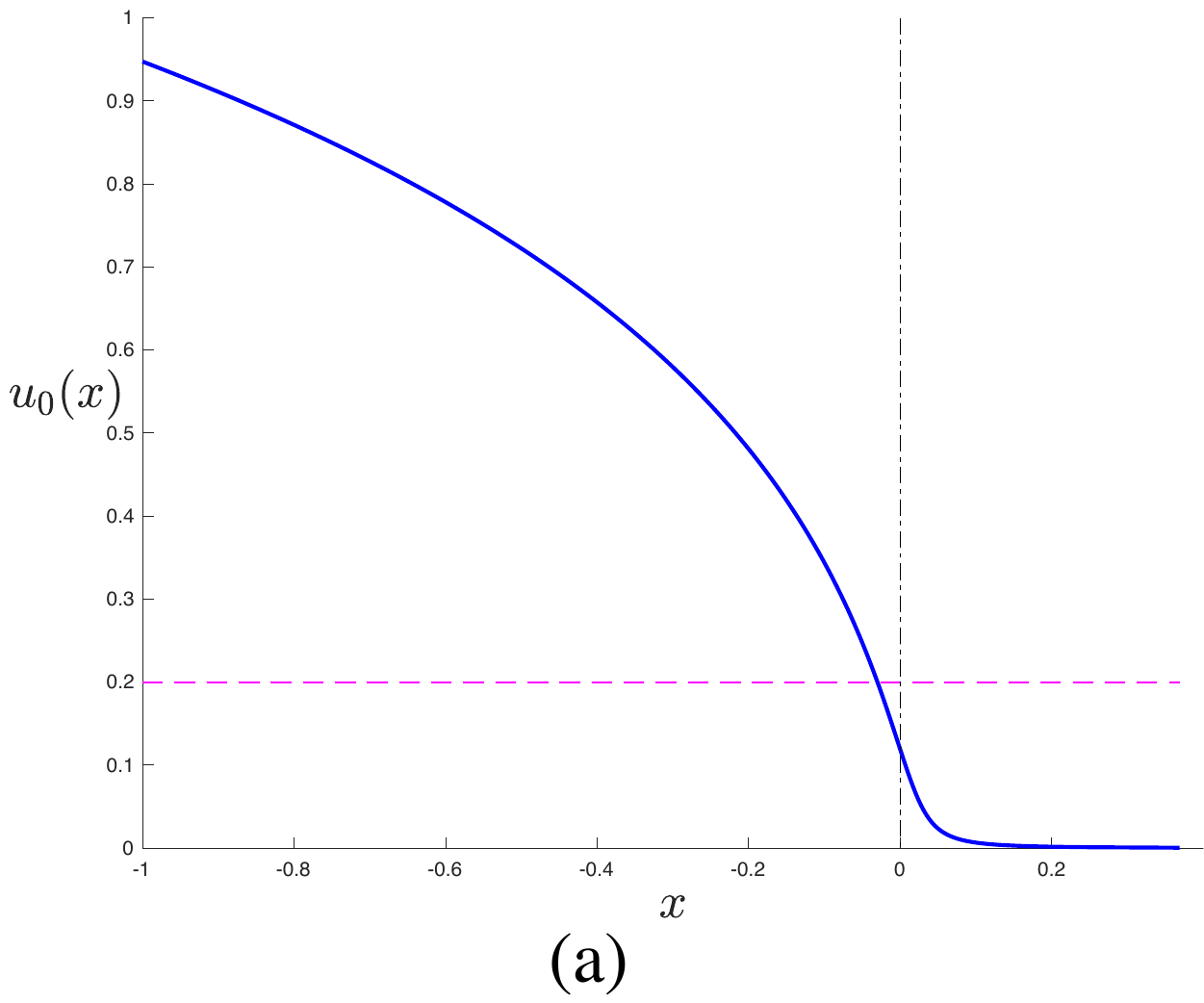}
    \includegraphics[width=0.49\textwidth]{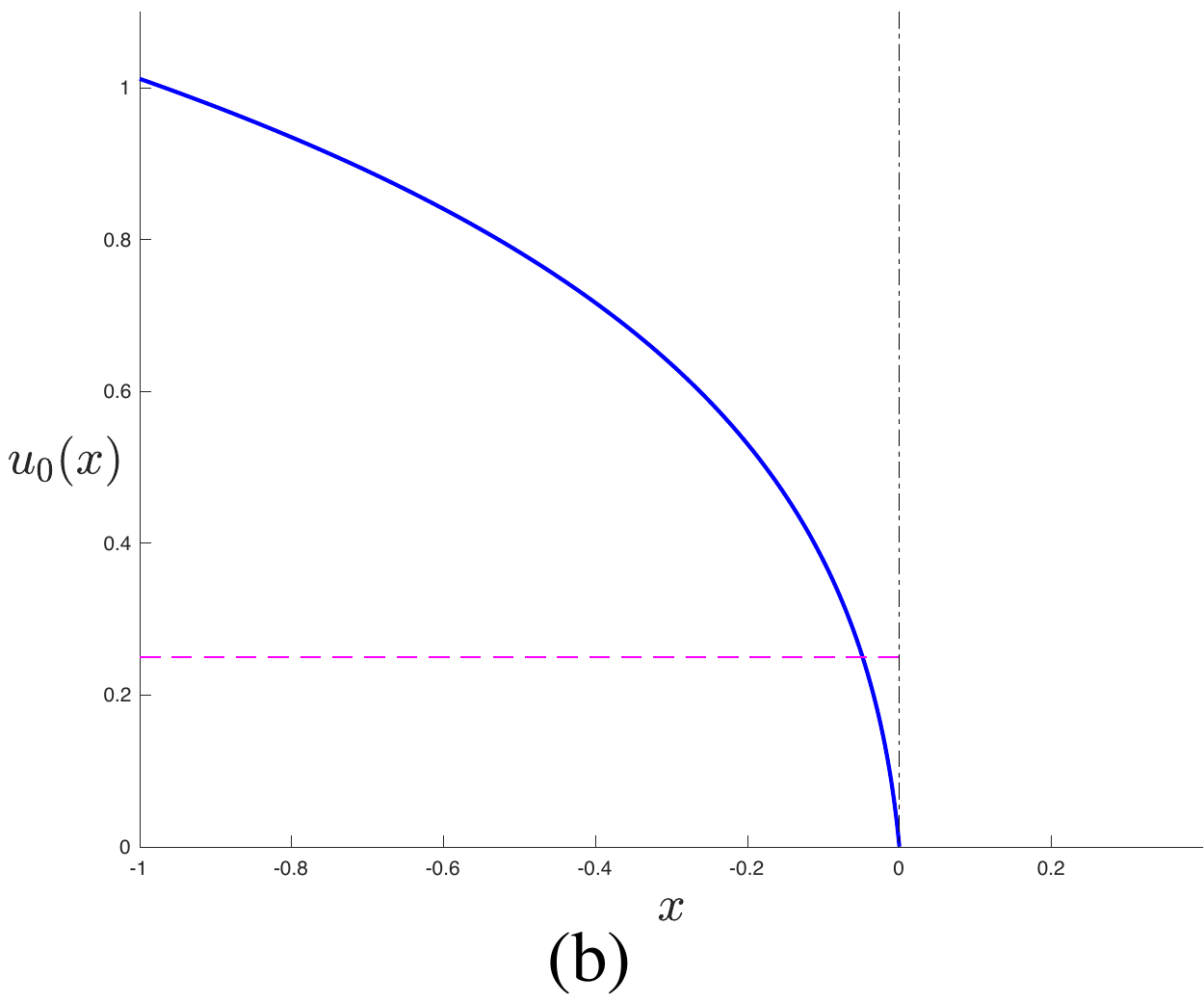}
    \caption{The solid curve is a plot of $u_0(x)$ for (a)  $E_0=0.2$, and  (b)  $E_0=0.25$, the values indicated by the horizontal dashed line. The sign of the derivative (and hence of the Jacobian $f(u_0)$) remains positive as long as  $E_0{\leq}\frac14$. Contrast with the cases presented in Figure~\ref{fig:u-plots-2}. (These curves are for the case of ${\cal N}{=}2$, with up to $k=7$ included from~(\ref{eq:teekay_Neq2}).)}
    \label{fig:u-plots-1}
\end{figure}

\begin{figure}[t]
    \centering
    \includegraphics[width=0.49\textwidth]{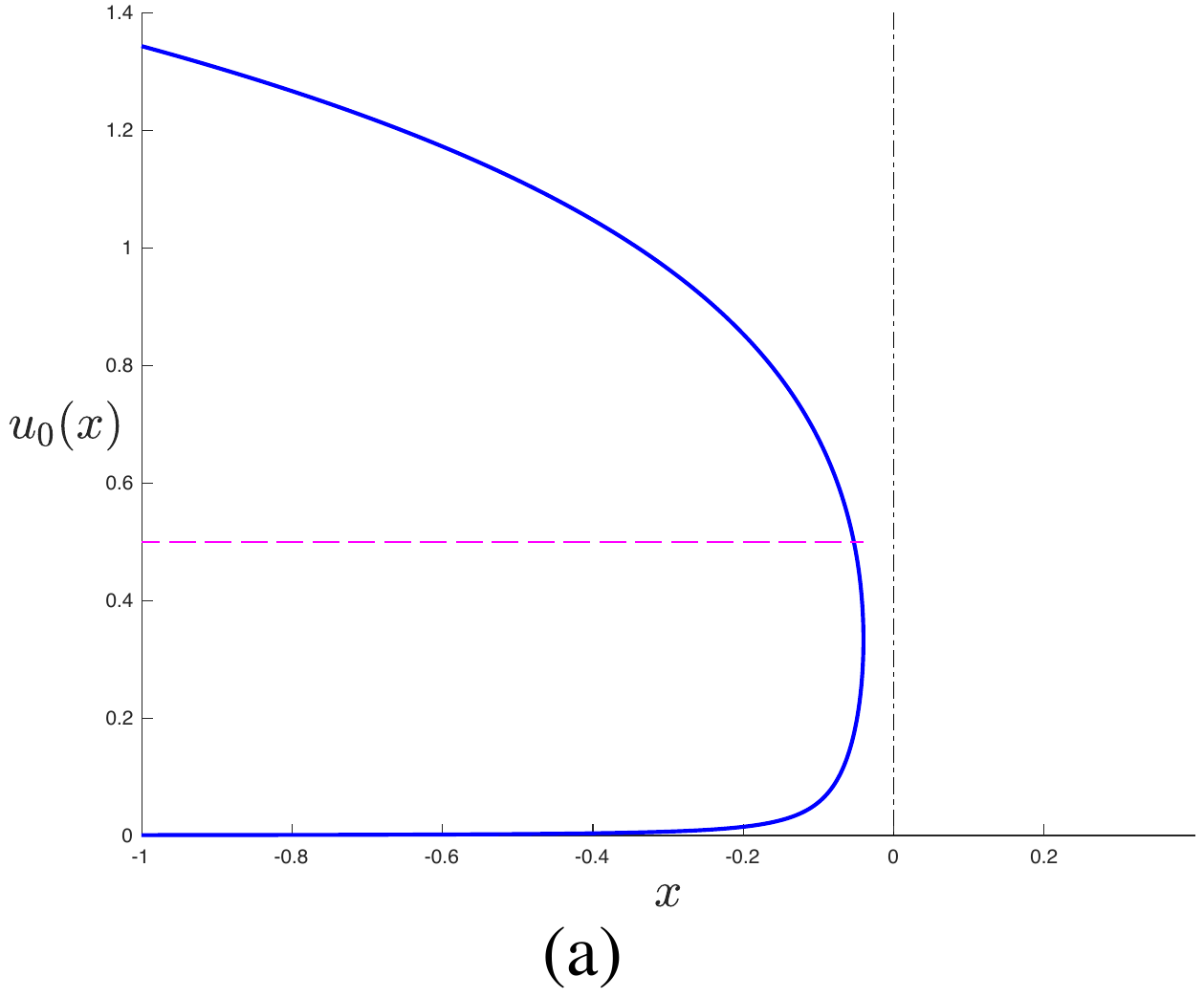}
    \includegraphics[width=0.49\textwidth]{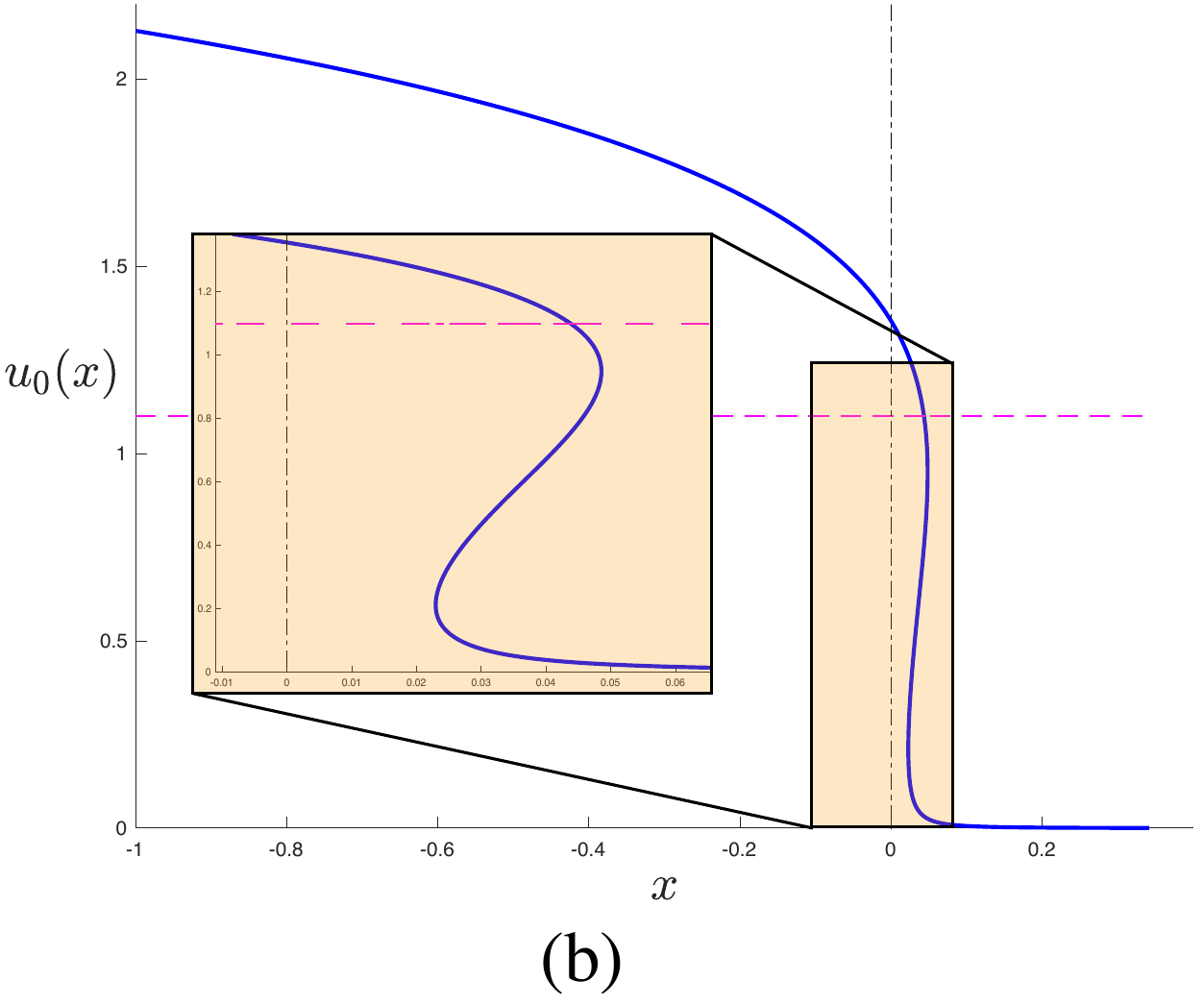}
    \caption{The solid curve is a plot of $u_0(x)$ for (a)  $E_0=0.5$, and  (b)  $E_0=1.1$, the values indicated by the horizontal dashed line. The sign of the derivative (and hence of the Jacobian $f(u_0)$) changes sign when $E_0{>}\frac14$, as can be seen in both cases. However, it remains positive for $u_0\in[E_0,\infty)$. (These curves are for the case of ${\cal N}{=}2$, with up to $k=7$ included from~(\ref{eq:teekay_Neq2}).)}
    \label{fig:u-plots-2}
\end{figure}
This leaves the analysis in a rather interesting position. It seems that (for $E_0 > \frac{1}{4}$) at the disc level, there is no globally-defined solution to the string equation that reproduces $\rho_0$. It is only  locally defined, but well enough (as we shall further emphasize in Section \ref{sec:exploration}) to compute all perturbative (in~$\hbar$) corrections to the partition function. 

However, the question of {\it non-perturbative} corrections is not so obvious\footnote{Certain avatars of this ``non-perturbative-instability'' were pointed out in \cite{Turiaci:2023jfa}. However, in their analysis they noticed problems only for multiplets that, in our conventions, correspond to the choice of minus sign in the $f$-function. The string equation on the other hand, reveals potential non-perturbative problems for all $E_0 > \frac{1}{4}$. We will discuss the connection between these observations in  Section~\ref{sec:exploration}.}. We will present initial explorations of that question in Section~\ref{sec:exploration} as well. For now, let us merely point out that for the theory at hand, the presence of BPS states is tantamount to $u_0 (x)$ being single-valued\footnote{This is because the condition $E_0<\frac{1}{4}$ translates to $|q|<\hat{q}$. The long multiplet of charge $q$ can be decomposed into two short multiplets with charge $q - \frac{\hat{q}}{2}$ and $q + \frac{\hat{q}}{2}$ and it is only one of them that has ground state degeneracy, provided $q - \textrm{sgn}(q) \frac{\hat{q}}{2}<\frac{\hat{q}}{2}$ and so these two conditions are indeed equivalent. One may wonder why only one short multiplet has non-zero degeneracy, since the long multiplet is built out of two short ones. It may be connected to the form (\ref{eq:0A-Q-matrix}) if one would interpret $M$ as acting on one short multiplet and $M^\dagger$ on another, since only one of matrices $MM^\dagger$ or $M^\dagger M$ could have generically large kernel. We postpone further investigations of this intuition to future work. We will see that in all other examples, it will be always one of the two short multiplets that has non-zero degeneracy.}, fitting nicely with criteria set out in ref.~\cite{Johnson:2021tnl}. We will see that this is in fact a leitmotif linking various examples described in this paper. It would be beneficial to get a better understanding of these connections and, preferably, be able to directly derive them from the full  matrix model. 

\section{Small ${\cal N}{=}4$ JT Supergravity}
\label{sec:small-Neq4-JT-supergravity}

The next demonstration is a revisit of ``small'' $\mathcal{N}{=}4$ JT supergravity. Before  diving in, we should explain the ``small/large''  terminology. The natural R-symmetry for  JT supergravity with $\mathcal{N}$-extended supersymmetry is $SO(\mathcal{N})$. However, one may encounter in practice theories in which the R-symmetry is smaller because certain generators act trivially on fields. In particular, if we dimensionally reduce the minimal $\mathcal{N}{=}2$ supergravity in 4D to AdS$_2$, we obtain $\mathcal{N}{=}4$ JT supergravity  with the R-symmetry group being $SU(2)$ \cite{Heydeman:2020hhw}. This theory is called {\it small} $\mathcal{N}{=}4$ JT supergravity and examples with more R-symmetries are called {\it large} $\mathcal{N}{=}4$ JT supergravity. The next section will be devoted to the latter.

The disc density of states at fixed angular momentum $J$ (an $SU(2)_R$ spin) reads~\cite{Heydeman:2020hhw}\footnote{\label{fn:factors-of-two}We are following here the conventions of \cite{Turiaci:2023jfa}. Earlier work \cite{Heydeman:2020hhw} contained a typo in normalization of the continuous piece of the spectral density, which propagated to ref.~\cite{Johnson:2024tgg}. We thank Joaquin Turiaci for useful discussions on this point.}:
\begin{equation}
 \rho_0(E,J) = \frac{1}{\pi \hbar} \frac{J \sinh \left(2\pi \sqrt{E-J^2}
    \right)}{\pi E^2} \Theta(E-J^2),
\end{equation}
where $J \in \frac{1}{2} \mathbb{Z}_{>0}$.
%
Here, $J^2$ plays the role of $E_0$ in the general framework laid out in Section~\ref{sec:new-approach}.  Rewriting $E {=} J^2 (1{+}\lambda)$, we get:
\begin{equation}
    \rho_0(\lambda, J) = \frac{1}{\pi \hbar} \frac{ \sinh \left( 2\pi J \sqrt{\lambda} \right)}{\pi J^3 (1+\lambda)^2}\ .
\end{equation}
At the first glance, this is a tragedy with respect to the prescription of  Section~\ref{sec:new-approach}. Here, $\rho_0$ has a double pole at $\lambda{=}{-}1$, something that  cannot be mapped to the matrix model. Not wishing to fall already at the second hurdle,  we inspect this pole more closely. It reads
\begin{equation}
    \rho_0(\lambda, J) =  \frac{i}{\pi \hbar} \frac{ \sin \left( 2\pi J \right)}{\pi J^3 (1+\lambda)^2} + O \left(\frac{1}{(1+\lambda)} \right)\ .
\end{equation}
We see that for physical values of the angular momenta $J \in \frac{1}{2} \mathbb{Z}$ the leading piece vanishes. We thus see that the requirement that the matrix model description exists imposes on us that the angular momentum is quantized! Again, we emphasize that this is not in any way  an input of the calculation; merely matching powers of $(1+\lambda)$ on both sides brings out this condition. Unfortunately, this condition renders our previous techniques useless, strictly speaking. We should not take $J$ to be discrete and then expand it around  zero\footnote{In fact, this strategy was successfully applied in ref.~\cite{Johnson:2024tgg}! This is because, fortunately, expanding preceded uncovering the discreteness. Progress was made by resumming an infinite number of problematic terms and showing that they cancel at discrete values of $J$. The prescription described in what follows, which yields the same answer, takes care of all these resummations automatically.}. There is however an elegant way out. Let us define:
\begin{equation}
    \widetilde\rho_0(\lambda, J) = \frac{1}{\pi \hbar} \frac{ \sinh \left( 2\pi J \sqrt{\lambda} \right)}{\pi J^3 (1+\lambda)^2} - \frac{\sqrt{\lambda}}{\pi \hbar} \frac{ \sin \left( 2\pi J \right)}{\pi J^3 (1+\lambda)^2}\ .
\end{equation}
The function $\widetilde\rho(\lambda, J)$ defined this way has two nice properties:
\begin{itemize}
    \item It has only a simple pole at $\lambda{=}{-}1$, so it admits a matrix model description {\it for any value of} $J$.
    \item For $J \in \frac{1}{2} \mathbb{Z}$ it reduces to $\rho(\lambda, J)$.
\end{itemize}
Hence, the next steps are clear. We define the matrix model with respect to $\widetilde\rho_0$ and then at the end of our calculations we simply impose $J \in \frac{1}{2} \mathbb{Z}$.
 (Although we will not need to use this fact, note that $\tilde\rho_0$ is positive definite for $J>0, \lambda>0$ so it is justified to interpret it as a density of states of some system.)

We are now ready to apply the general scheme of Section~\ref{sec:new-approach} (writing $E_0 = J^2$). For $\widetilde\Gamma$ we obtain the condition that was arrived at in ref.~\cite{Johnson:2024tgg}:\footnote{See footnote~\ref{fn:factors-of-two} concerning the factor of 2 conversion.}
\begin{equation}
   \pm |\widetilde\Gamma| = -2 \cos (2\pi J) + \frac{\sin (2\pi J)}{ \pi J}\ .
\end{equation}
For physical values of $J$, it follows that $\widetilde\Gamma = 2$ and we choose plus for $J \in \mathbb{N}+\frac{1}{2}$ and minus for $J \in \mathbb{N}_{\ge 1}$. One may be worried that $\widetilde{\Gamma} =2$, whereas the only BPS states are in $J=0$ sector and they have $\widetilde{\Gamma}=1$. To see the source of this (apparent) discrepancy, we should go over the derivation of densities for this theory \cite{Heydeman:2020hhw}. One starts with the sum over all saddles (together with their one-loop determinants) and, using Poisson resummation, rewrites it as a sum over representations of the $R$ symmetry. These are labelled by half-integers. However, each spin, {\it except} for $j=0$ appears in this sum twice. This is the source of this (apparent) mismatch. It is not a mismatch at all, but a result of the convention for how the states are counted.

The next step is to move the term with $\widetilde\Gamma$ to the left:
\begin{equation}
    {\rm LHS} = \frac{1}{\pi \hbar} \frac{ \sinh \left( 2\pi J \sqrt{\lambda} \right)}{\pi J^3 (1+\lambda)^2} - \frac{\sqrt{\lambda}}{\pi \hbar} \frac{ \sin \left( 2\pi J \right)}{\pi J^3 (1+\lambda)^2} - \frac{1}{\pi \hbar} \frac{\sqrt{\lambda}}{J^2(1+\lambda)} \left(
    - \cos (2\pi J) + \frac{\sin (2\pi J)}{2 \pi J}\right)\ .
\end{equation}
Now, we may simply expand it in powers of $J$ and compare with $t_k$s on the right hand side. It should be clear from the construction that at $n$th order in $J^2$, the answer will be of the form $\sqrt{\lambda}$ times a polynomial in $\lambda$ of degree $n-1$. Indeed, it expands to: 
\begin{equation}
   {\rm LHS} =  \sum_{k=1} 2 J^{2k} \frac{(-4)^{k+1} \pi ^{2 k+1} \sqrt{\lambda }}{\hbar (2k+3)!} \times \sum_{n=0}^{k-1}(-\lambda)^n (k-n)\ ,
\end{equation}
and this should be equal to:
\begin{equation}
   {\rm RHS} =  \frac{1}{2\pi \hbar} \sum_{k=1}t_k k J^{2k-1} \sqrt{\lambda} w_k (\lambda)\ ,
\end{equation}
where $w_k(\lambda)$ was defined in~(\ref{eq:w_k-lambda}). So $t_k$ must have the following expansion:
\begin{equation}
    t_k = J \sum_{n=0} t_{k,n}J^{2n}\ ,
\end{equation}
giving:
\begin{equation}
   {\rm RHS} =  \frac{\sqrt{\lambda}}{2\pi \hbar} \sum_{k=1}  J^{2k} \sum_{n=0}^{k-1} (k-n) t_{k-n,n} w_{k-n}(\lambda)\ .
\end{equation}
Comparing factors of $J$ on both sides, we get:
\begin{equation}
    4 \frac{(-4)^{k+1} \pi^{2k+2}}{(2k+3)!} \sum_{n=0}^{k-1} (-\lambda)^n (k-n) = \sum_{n=0}^{k-1} (k-n) t_{k-n,n} w_{k-n}(\lambda)\ .
\end{equation}
Solving it step by step we find:
\begin{eqnarray}
    t_{1,0} &=& \frac{4}{15} \pi^4
\ ,\quad
    t_{1,1} = - \frac{4}{45} \pi^6
\ ,\quad
    t_{1,2} = \frac{1}{90} \pi^8
\ ,\quad
    t_{1,3} = -\frac{1}{1350} \pi^{10}
\ ,\cdots\quad
\nonumber \\
\Longrightarrow    t_{1,n}&=& \frac{8}{15} \frac{(-1)^n}{n!(n+2)!} \pi^{2n+4} \nonumber \\
    t_{2,0} &=& \frac{2}{105} \pi^6
\ ,\quad
    t_{2,1} = - \frac{1}{210} \pi^8
\ ,\quad
    t_{2,2} = \frac{1}{2100} \pi^{10}
\ ,\quad
    t_{2,3} = - \frac{1}{37800} \pi^{12}
\ ,\cdots\quad
\nonumber \\
\Longrightarrow     t_{2,n} &=& \frac{4}{35} \frac{(-1)^n}{n! (n+3)!} \pi^{2n+6}\ ,
\nonumber\\
    t_{3,0} &=& \frac{1}{1134}\pi^8
\ ,\quad
    t_{3,1} = -\frac{1}{5670} \pi^{10}
\ ,\quad
    t_{3,2} = \frac{1}{68040} \pi^{12}
\ ,\quad
    t_{3,3} = -\frac{1}{1428840} \pi ^{14}
\ ,\cdots\quad
\nonumber \\
\Longrightarrow     t_{3,n} &=& \frac{4}{189} \frac{(-1)^n}{n!(n+4)!} \pi^{2n+8}
\ ,\nonumber \\
\Longrightarrow     t_{k,n} &=& 8\frac{1}{k!(2k+3)(2k+1)} \frac{(-1)^n}{n! (n+k+1)!} \pi^{2+2k+2n}\ .
\end{eqnarray}
This can be summed to yield:
\begin{equation}
    t_k = \frac{J 2^{k+4} \pi^{2k+2}}{k! (2k+1)(2k+3)} \frac{J_{k+1}(2\pi J)}{(2\pi J)^{k+1}}\ .
\end{equation}
This answer is consistent with the answer found in ref.~\cite{Johnson:2024tgg} (up to the overall aforementioned difference of a factor of 2 in  the normalization of $\rho_0(E)$).

Equipped with the explicit solutions, we may study them in detail. The first observation one can make is about the sign in front of $\widetilde{\Gamma}$ in our solution. It is plus for $J \in \mathbb{N}+\frac{1}{2}$ and minus for $J \in \mathbb{N}_{\ge 1}$. Thus, it follows, by the same argument as for  $\mathcal{N}{=}2$, that the latter must have multivalued~$u_0 (x)$. Moreover, one can check that for $J \in \mathbb{N}_{\ge 1}+\frac{1}{2}$, $u_0(x)$ fails to be single-valued as well. Thus, the only sector that avoids these issues is $J=\frac{1}{2}$. Note that the long multiplet with $J=\frac{1}{2}$ decomposes into short multiplets with $J=0$ and $J=\frac{1}{2}$. All these observations are perfectly consistent with our conjectures connecting single-valuedness and the presence of BPS states, since the latter are present only at $J=0$ for the theory at hand. Let us finish by noting that in all examples, the signs of multi-valuedness appear for $u_0<E_0$ and as such do not affect the  perturbative (in $\hbar$) description. Section~\ref{sec:perturbative-remarks} discusses this further.

Having whetted our tools on the two familiar examples,   it it time to turn to new JT supergravity models, recently published in ref~\cite{Heydeman:2025vcc}. We will show  that, with care, they  fit the requirement of being captured as a multicritical matrix model, including  {\it all  the details of  the correct BPS spectrum}.\footnote{We thank Joaquin Turiaci for sharing  preliminary results with us at an early stage. This allowed us to even successfully test out the {\it predictive} power of our methods, computing details of the BPS sector in some models before we received the final answer.}

\section{Large ${\cal N}{=}4$ JT Supergravity}
\label{sec:large-Neq4-JT-supergravity}

This theory was presented in ref.~\cite{Heydeman:2025vcc}. Charged sectors are described by two non-negative integers:~$j_+$,~$j_-$. The density of states was found to be:
\begin{subequations}
\begin{equation}
       \widetilde{\rho}_{j_+, j_-} = \frac{1}{2\pi \hbar} A_{j_+, j_-} \frac{\sinh \left(
2\pi \sqrt{E-E_0}
    \right)}{\left(
E-E^-
    \right) \left(E-E^+ \right)} \Theta \left(
    E-E_0
    \right)\ ,
\end{equation}
where:
\begin{equation}
    A_{j_+, j_-} =  \frac{\alpha^{3/2} j_{+} j_{-} \sqrt{2\pi}}{16 (1+\alpha)^3}\ ,\quad
    E^\pm = \frac{\alpha (j_+ \pm j_-)^2}{(1+\alpha)^2}\ ,\quad
    {\rm and}
    \quad
    E_0 = \frac{\alpha j_+^2 + j_-^2}{(1+\alpha)}\ .
\end{equation}
\end{subequations}
It is easy to see that $E_0 \ge E^\pm$ and so $\widetilde{\rho}_{j_+, j_-}$  remains positive, as discussed in detail in ref.~\cite{Heydeman:2025vcc}. 

As it stands, this density of states is problematic from the point of view of the random matrix model. We argued that (within the considered class of matrices), the spectral density can have at most one pole, but $\widetilde{\rho}_{j_+,j_-}$ has two!\footnote{With the obvious exception of the case $j_+ j_- = 0$ but then also $A_{j_+ j_-} = 0$.} 

This  leads us to a certain modification. Note that:
\begin{equation}
    \widetilde{\rho}_{j_+, j_-} = \frac{1}{2\pi \hbar} \frac{\sqrt{2\pi \alpha}}{64 (1+\alpha)} \left(
\frac{1}{E-E^+} - \frac{1}{E-E^-}
    \right) \sinh \left(
2\pi \sqrt{E-E_0}
    \right) \Theta \left(
    E-E_0
    \right).
\end{equation}
Motivated by this decomposition, we propose new spectral densities:
\begin{equation}
    \rho_{j_+,j_-}^\pm = \frac{1}{2\pi \hbar} \frac{\sqrt{2\pi \alpha}}{64 \pi (1+\alpha)}\frac{1}{E-E^\pm}\sinh \left(
2\pi \sqrt{E-E_0}
    \right) \Theta \left(E-E_0 \right),
\end{equation}
such that:
\begin{equation}
\widetilde{\rho}_{j_+,j_-} = \rho_{j_+,j_-}^+ - \rho_{j_+,j_-}^-\ ,
\end{equation}
and $\rho_{j_+,j_-}^\pm$ originate from two statistically independent matrix models.
At first glance, it may seem arbitrary. However, if one carefully studies the derivation of $\widetilde{\rho}_{j_+,j_-}$ (as presented in Appendix~A of ref.~\cite{Heydeman:2025vcc}; see especially (A.12)), one can see that it arises as a sum over sectors with different angular momenta, both positive and negative\footnote{To be more precise, for given absolute values of $j_+,j_-$ there are four possible terms, will all possible signs. We decided to sum $++$ and $--$ together and also $+-$ and $-+$. Other possibilities are not excluded but they necessarily lead to complex spectral densities.}. Here, we simply decided to keep terms with $j_+ j_- $ positive and negative separate or, in other words, we postulate that these are two statistically-independent sectors\footnote{This could be checked by careful computation of the cylinder amplitude. As such, one may treat our proposal as a testable prediction of the string equation. A conservative way to interpret this result is that, since the matrix model encodes the topological recursion of the gravitational path integral, the string equation informs us about the non-trivial structure of said recursion. See ref.~\cite{Johnson_upcoming} for some upcoming work on using the string equation, combined with the Gel'fand-Dikii equation, to explore the (super) Weil-Peterson volumes that result from extended JT supergravity.}. One may be also worried by the minus sign in front of $\rho^{-}_{j_+,j_-}$, but since the spectral densities are from separate ensembles,  it is meaningful (in a probabilistic sense) to combine them in this way. More detailed gravitational path integral computations could shed more light on the interpretation of this sign -- we postpone it for future work. It is possible that the sign is connected to the non-linear deformation of the superalgebra that appears in the theory\footnote{We thank Joaquin Turiaci for this suggestion.}.

A nice thing about $\rho_{j_+, j_-}^\pm$ is that this is essentially $\mathcal{N}{=}2$ JT supergravity, only with the shifted (and rescaled) spectral density. Implementation of the shift of the spectral density amounts to simply shifting random matrix $H {=} M^\dagger M$ by a constant (times $N \times N$ identity matrix). It leads to the same string equation but with a shifted $u$. Thus, we may immediately read off all the coefficients:
\begin{subequations}
    \begin{equation}
|\widetilde{\Gamma}^\pm| = \pm \frac{\sqrt{2\pi \alpha}}{64\pi (1+\alpha)} \sin \left(
        2\pi \sqrt{E_0 - E^\pm}
        \right)\ ,
    \end{equation}
        \begin{equation}
    t_k^\pm =  \frac{\sqrt{2\pi \alpha}}{64\pi (1+\alpha)} \frac{2\pi ^{k+1} \left(E_0 - E^\pm \right)^{-\frac{k}{2}} J_k\left(2 \pi\sqrt{E_0 - E^\pm}  \right)}{k!(2k+1)}\ ,
\end{equation}
\end{subequations}
Notice that:
\begin{subequations}
    \begin{equation}
  \biggl| \sin \left(2\pi \sqrt{E_0 - E^+} \right)\biggr| =\biggl| \sin \left( 2\pi\frac{\alpha j_+ - j_-}{1+\alpha} \right)\biggr| = \biggl|\sin \left( 
\frac{2\pi \alpha(j_+ + j_-)}{1+\alpha} -2\pi j_-
  \right)\biggr| = \biggl|\sin \left( 
\frac{2\pi \alpha(j_+ + j_-)}{1+\alpha}
  \right)\biggr|\ .
\end{equation}
The last equality follows from $j_- \in \frac{1}{2} \mathbb{Z}$ and so:
\begin{equation}
    |\widetilde{\Gamma}^+| = \hbar^{-1}N_{j_+,j_- -1/2} = \hbar^{-1}N_{j_+ - 1/2,j_-}\ ,
\end{equation}
\end{subequations}
where $N_{j_+,j_-}$ is, according to ref.~\cite{Heydeman:2025vcc} the number of BPS states in the short multiplet $(j_+,j_-)$. Moreover, we would expect the BPS energy to corresponds to the pole so:
\begin{equation}
    E_{{\rm BPS}, j_+ - 1/2, j_-}=E_{{\rm BPS}, j_+, j_- -1/2} = E^+ = \frac{\alpha}{(1+\alpha)^2} (j_+ + j_-)^2\ ,
\end{equation}
which again agrees with ref.~\cite{Heydeman:2025vcc}. Notice that the degenerate states could contribute both to $(j_+,j_- {-} 1/2)$ and $(j_+ {-} 1/2,j_-)$ short multiplets, so it is natural to ask which it is.\footnote{In general, it could also be $(j_+ {+} k, j_- {-} k {-}1/2)$ or $(j_+ {+} k{-}1/2, j_- {-}k)$ for $k \in \frac{1}{2}\mathbb{Z}$, but we will see these multiplets do not appear.} To answer this question, we must remember one more fact about $\mathcal{N}{=}2$, namely that these matrix models enjoy BPS states only if their would-be-gap is not too large: it must be smaller than one-fourth (so that its square root must be smaller than one-half). The would-be-gap in our case is $\left(\frac{\alpha j_+ - j_-}{1+\alpha} \right)^2$. Depending on the sign of the expression in the bracket, we will have two cases. If it is positive, we have:
\begin{equation}
    \frac{1}{2} > \frac{\alpha j_+ - j_-}{1+\alpha} = \frac{\alpha}{1+\alpha} \left(
\left(j_+-\frac{1}{2} \right)+ j_- + \frac{1}{2}
    \right) - j_- > 0 \ .
\end{equation}
This is the condition for $(j_+ - \frac{1}{2}, j_-)$ short multiplet to have BPS states \cite{Heydeman:2025vcc}. If, on the other hand, it is negative, we can write:
\begin{equation}
    0 > \frac{\alpha j_+ - j_-}{1+\alpha} = \frac{\alpha}{1+\alpha} \left(
j_+ + \left(j_- - \frac{1}{2} \right) + \frac{1}{2}
    \right) - \left(j_- - \frac{1}{2} \right) - \frac{1}{2} > 0\ .
\end{equation}
This is the condition for $(j_+, j_- - \frac{1}{2})$ short multiplet to have BPS states. The appearance of $(j_+ - 1/2, j_-)$ and $(j_+, j_- - \frac{1}{2})$ is not accidental. Long multiplet $(j_+, j_-)$ is in fact a direct sum of these two! 

What about the minus sector, does it also know about BPS states? The would-be gap reads:
\begin{equation}
    \frac{\alpha j_+ + j_+}{1+\alpha} \ge \textrm{min}(j_+, j_-) \ge \frac{1}{2}\ ,
\end{equation}
and thus it follows that there are no BPS states in the minus sector.\footnote{Interestingly, if we ignored the fact that the gap cannot be too large we would find non-zero number $N$ of states located at $E{=}{E^-}$. Both the number $N$ and the energy ${E^-}$ can still be found in \cite{Heydeman:2025vcc}. Indeed, the expression for the BPS density of states from this paper contains a $\Theta$-function written in terms of $j_+, j_-$, which exactly implements that the gap cannot be too large. Had we ignored this $\Theta$-function, we would find both that there are additional $\delta$-functions located at $E{=}{E^-}$ and that coefficient is exactly $N$. Thus, it seems that the matrix model not only knows about non-zero BPS contributions but it also correctly predicts the functional form of the zero contributions.}

\section{${\cal N}{=}3$ JT Supergravity}
\label{sec:Neq3-JT-supergravity}
This model was first introduced in ref.~\cite{Heydeman:2025vcc}. In fact, it admits three (possible) realizations (labeled by the R-symmetry): $SO(3)$, anomalous $SO(3)$ and $SU(2)$. The states in these theories transform in spin-$j$ representations with $j \in \mathbb{N}, \mathbb{N}+\frac{1}{2}, \frac{1}{2} \mathbb{N}$, respectively. Currently, we are only able to construct a multicritical matrix model description for the anomalous case\footnote{The other two theories have analytic structure which does not seem to be consistent with the string equation. It may be connected to the fact that the anticommutator of supercharges is non-linearly corrected, as noted in ref.~\cite{Heydeman:2025vcc}.  This could also be related to the fact that density of states in large $\mathcal{N}{=}4$ theory is also not consistent with the string equation if one does not perform the conjectured split into two statistically independent matrix models.} so this  will be our focus from now on, although see the end of this section for some remarks about other matrix model realizations.

The density of states reads:
\begin{subequations}
\begin{equation}
        \rho_j(E) = \frac{1}{2\pi \hbar} \frac{j \cosh \left(
        2\pi \sqrt{E-E_0(j)}
        \right)}{2 E \sqrt{E-E_0(j)}}\ ,
        \label{eq:Neq3-denisty}
\end{equation}
where $j \in \mathbb{N}+\frac{1}{2}$ and
\begin{equation}
    E_0(j) = \frac{j^2}{4}\ .
\end{equation}
\end{subequations}
Writing: 
\begin{equation}
    \frac{1}{E \sqrt{E-E_0}} = \frac{1}{E_0} \left(
\frac{1}{\sqrt{E-E_0}} - \frac{\sqrt{E-E_0}}{E}
    \right)\ ,
    \label{eq:splitting}
\end{equation}
it is clear  that this model is structurally a non-trivial combination of $\mathcal{N}{=}1$ and $\mathcal{N}{=}2$ JT supergravity. At  first glance, this seems to be a disaster, since this model has both a pole at $E=0$ (the $\mathcal{N}{=}2$ piece) and the $(E-E_0)^{-1/2}$ edge (the $\mathcal{N}{=}1$ piece) instead of the usual $(E-E_0)^{1/2}$. However, this tension is illusive since as $E \to 0$, $\cosh \left(
2\pi \sqrt{E-E_0}
\right) \to \cos(\pi j) = 0$, since $j \in \mathbb{N}+\frac{1}{2}$. Thus, there is no pole at~$E{=}0$ (and, importantly, nowhere else). We must now deal with the $(E-E_0)^{-1/2}$ edge. This is the classic hard edge behaviour discussed in Section~\ref{sec:N_eq_1_supergravity}, seen earlier in $\mathcal{N}{=}1$ JT supergravity, albeit with~$E_0 {=}0$. To incorporate the shift, we must consider the slightly more general string equation given in equation~(\ref{eq:slightly-bigger-string-equation}), setting $\sigma=E_0$ and taking the classical limit as before, giving ({\it c.f.,} equation~\ref{eq:leading-Neq-1-string-equation}):
\begin{equation}
    \left(u_0-E_0 \right) \mathcal{R}_0^2 = 0\ ,\quad{\rm with}\quad {\cal R}_0[u_0]\equiv\sum_{k=1}^\infty t_k u_0^k+x\ .
\end{equation}
As before,  this equation has two solution branches, corresponding to:
\begin{eqnarray}
        u_0 &=& E_0\ , \quad {\rm for} \quad x>b\ ,\quad {\rm with} 
    \nonumber\\
        \mathcal{R}_0(x,u_0) &=& 0 \ ,
        \quad {\rm for} \quad x\leq b \ , 
\end{eqnarray}
where $b$ ensures a (piecewise) continuous solution, and is to be determined.
We may build the full solution from these two branches. Note that our formula for the leading density of states \eqref{eq:leading-representation-u0} is valid only in the regime in which $f = -\frac{dx}{du_0}$ is nonzero. In particular, this excludes the first branch for which we must use instead \eqref{eq:leading-representation}. Then, we see that:
\begin{equation}
    \rho_0(E) = \frac{1}{2\pi \hbar} \sum_{k=1}t_k k E_0^{k-1/2} \sqrt{\lambda} w_k (\lambda)   + \frac{(\mu-b)}{2\pi \hbar E_0^{1/2}} \frac{1}{\sqrt{\lambda}}\ . 
\end{equation}
Parameter $\mu-b$ is fixed by comparing coefficients in front of $\lambda^{-1/2}$. Then, all the $t_k$s are determined as in other cases. Explicitly, we find: 
\begin{subequations}
\begin{equation}
    \rho_j(\lambda) = \frac{1}{2\pi \hbar} \frac{\cosh \left(2 \pi \sqrt{E_0 \lambda} \right)}{E_0} \left(
\frac{1}{\sqrt{\lambda}} - \frac{\sqrt{\lambda}}{1+\lambda}
    \right)\ , \label{N3_split}
\end{equation}
and so:
\begin{equation}
    \mu-b = \frac{1}{E_0^{1/2}}\ .
\end{equation}
\end{subequations}
To determine $t_k$s, we may introduce an auxiliary density:
\begin{equation}
    2\pi \hbar \tilde{\rho}_j(\lambda) = \frac{\cosh \left(2 \pi \sqrt{E_0 \lambda} \right)}{E_0} \left(
\frac{1}{\sqrt{\lambda}} - \frac{\sqrt{\lambda}}{1+\lambda}
    \right) - \frac{1}{E_0 \sqrt{\lambda}} + \frac{\sqrt{\lambda}}{E_0 (1+\lambda)} \cos \left(2\pi \sqrt{E_0} \right)\ ,
    \label{eq:Neq3-auxiliary}
\end{equation}
which has the same set of coefficients $t_k$s (when $j \in \mathbb{N} + \frac{1}{2}$) but is free of all singularities (except for branch cuts) for all $j$. It is easy to find a Taylor expansion in small $E_0$:
\begin{equation}
    2\pi \hbar \tilde{\rho}_j(\lambda) = \sum_{n=1} \sqrt{\lambda} \frac{(2\pi)^{2n+2}}{(2n+2)!} \sum_{k=0}^{n-1} (-\lambda)^k (-1)^{n+1} E_0^n\ .
\end{equation}
We see in particular that this expansion has only integer and positive powers of $E_0$. It follows that $t_k$s must have only half-integer powers of $E_0$. Thus, let us slightly modify our previous ansatz:
\begin{equation}
    t_k = \sqrt{E_0} \sum_{n=0} t_{k,n} E_0^n\ .
\end{equation}
Explicitly, we obtain:
\begin{equation}
    \sum_{k=1} t_k k E_0^{k-1/2} w_k(\lambda) = \sum_{n=1} E_0^n \sum_{k=0}^{n-1} (n-k) t_{n-k,k} w_{n-k}(\lambda)\ .
\end{equation}
Comparing coefficients in front of powers of $E_0$, we get:
\begin{equation}
    (-1)^{n+1} \frac{(2\pi)^{2n+2}}{(2n+2)!} \sum_{k=0}^{n-1} (-\lambda)^k = \sum_{k=0}^{n-1} (n-k) t_{n-k,k} w_{n-k}(\lambda)\ .
\end{equation}
Note that the right hand side is almost the same as for $\mathcal{N}{=}2$. Thus, we immediately see that:
\begin{equation}
    t_{n-k,k} = \frac{(2\pi)^3}{2n+2} t_{n-k,k}^{\mathcal{N}{=}2} \ ,\quad \Longrightarrow\quad
    t_{k,n} = \frac{2 (-1)^n \pi ^{2 (k+n+1)}}{(2 k+1) k! n! (k+n+1)!}\ .
\end{equation}
This can be easily summed to yield:
\begin{equation}
    t_k = \frac{2 \pi ^{k+1} J_{k+1}\left(2 \pi  \sqrt{E_0}\right)}{(2 k+1) E_0^{k/2} k!}\ .
\end{equation}
In particular, we have
\begin{equation}
    t_k^{\mathcal{N}{=}4} = \frac{2}{k+3/2} t_k^{\mathcal{N}{=}3}\ .
\end{equation}
The fact that this model has no BPS sector means that it could also be prudent to seek a realization in the 0B framework discussed in the review of section~\ref{sec:string-equations-general}. As a reminder, it is an Hermitian matrix model for the supercharge $Q$, described as a simple multicritical merging two-cut system. The natural (scaled) $Q$-eigenvalues, denoted~$q$, would be on the real line, and a (shifted) relation to those of $H$ would be $q^2{=}E{-}E_0$. Changing  variables in the density~(\ref{eq:Neq3-denisty}) to $\rho_0(q){=}j\cosh(2\pi q)/(4\pi\hbar(q^2+E_0))$, there is potentially  a pair of poles on the imaginary axis at $\pm iE_0$, but their coefficients vanish for this case, as before (see below~(\ref{eq:splitting})). This then leaves the way clear to follow the procedures done above (starting with~(\ref{eq:Neq3-auxiliary})), now in the $q$ variable, to determine the $t_k$s.
In this way, we see that there is indeed a (perhaps natural) home for one of the ${\cal N}{=}3$ models in 0B.

Encouraged by this, one might try to see if all the ${\cal N}{=}3$ models might have a home in 0B.\footnote{We thank Joaquin Turiaci for suggesting this.} Looking back at the (non-anomalous) cases mentioned at the beginning of this section, it is harder to know how to proceed, as then (just as in the 0A framework) it is unclear what to make of the pair of poles along the complex $q$ axis. They lie outside the framework developed in Section~\ref{sec:new-approach}, coming from combining the leading string equation with the integral representation of the density. (At leading order the difference in the analysis for the 0B variables amounts to a simple change of variables.) Furthermore, a preliminary analysis suggests that turning on the parameters  described in refs.~\cite{Crnkovic:1990mr,Crnkovic:1992wd} do not seem to generalize the string equations in the right way to incorporate the BPS sector. However, it  seems that a large parameter space of options might be available, perhaps corresponding to coupling the $Q$ random matrix model to an appropriate fixed matrix, which seems to give new classes of behaviour. (See {\it e.g.,} refs.~\cite{Gross:1991aj,bleher2010randommatrixmodelexternal,orantin2008gaussianmatrixmodelexternal} for some examples.) This seems like an interesting avenue of investigation, but it is beyond the scope of this paper.

Notice that there is something else that could be done in the 0A framework. Let's go back to \eqref{N3_split} and compare it with what we did in  Section~\ref{sec:large-Neq4-JT-supergravity}. For large $\mathcal{N}{=}4$ JT supergravity, we decided to perform an analogous split and then conjecture that the two terms correspond to statistically independent matrix models. So why did we follow a different strategy for $\mathcal{N}{=}3$? The reason is that the derivation of the former theory incorporated in a very natural way these two sectors (they corresponded to different terms in the Poisson-resummed partition function) and, as far as we are aware of, there is no such structure for $\mathcal{N}{=}3$. In particular, it made the split canonical---otherwise we could always move a factor from one sector to another. Nevertheless, we cannot exclude the possibility that something similar is at play here. For the sake of completeness, we will work out the matrix models corresponding to \eqref{N3_split}, for all the ${\cal N}{=}3$ variants. The first component of the split  is:
\begin{equation}
    \rho^{(1)} = \frac{1}{2\pi \hbar} \frac{\cosh \left(2 \pi \sqrt{E_0 \lambda} \right)}{E_0 \sqrt{\lambda}}\ .
\end{equation}
This is very clearly structurally an  $\mathcal{N}{=}1$ model, albeit shifted and rescaled. Thus, it is immediate to see that the leading order string equation must read
\begin{equation}
    I_0 \left(2\pi \sqrt{u_0 - E_0} \right) -1 + E_0 x =0 \ .
\end{equation}
The second component is:
\begin{equation}
    \rho^{(2)} = \frac{1}{2\pi \hbar} \frac{\sqrt{\lambda} \cosh \left(
    2\pi \sqrt{E_0 \lambda}
    \right)}{E_0 \left(1+\lambda \right)}\ .
\end{equation}
We can read off:
\begin{equation}
    |\widetilde{\Gamma} |= \pm \frac{\cos \left(2\pi \sqrt{E_0} \right)}{E_0}\ .
\end{equation}
Moving this pole to the left, we are left with:
\begin{equation}
   {\rm LHS} =  \frac{1}{E_0 (1+\lambda)} \left(
\cosh \left(2\pi \sqrt{E_0 \lambda} \right) - \cos\left(2\pi \sqrt{E_0} \right)
    \right) = \sum_{n=1} \frac{(2\pi)^{2n}}{(2n)!} (-E_0)^{n-1} \sum_{k=0}^{n-1} (-\lambda)^k\ .
\end{equation}
We see that this is a series in integer powers of $E_0$, starting from zero. We are thus lead to write
\begin{equation}
    t_k(E_0) = E_0^{-1/2} \sum_{n=0} t_{k,n}E_0^n\ ,
\end{equation}
so that:
\begin{equation}
    {\rm RHS} = \sum_{k=1} k t_k E_0^{k-1/2} w_k (\lambda)  = \sum_{k=1} \sum_{n=0} k t_{k,n} E^{n+k-1} w_k (\lambda) = \sum_{n=1}  E_0^{n-1} \sum_{k=0}^{n-1} (n-k) t_{n-k,k} w_{n-k}(\lambda)\ .
\end{equation}
Comparing different powers of $E_0$, we find:
\begin{equation}
    \frac{(2\pi)^{2n}}{(2n)!} (-1)^{n-1} \sum_{k=0}^{n-1} (-\lambda)^k = \sum_{k=0}^{n-1} (n-k) t_{n-k,k} w_{n-k}(\lambda)\ .
\end{equation}
The right hand side is of course the same as for $\mathcal{N}{=}2$, so we can immediately read off:
\begin{equation}
    t_{n-k,k}^{\mathcal{N}{=}3} = 2\pi (2n+1) t_{n-k,k}^{\mathcal{N}{=}2}  = \frac{(-1)^k (2n+1) \pi^{2n}}{k! n! (n-k)! (2n-2k +1)}\ ,
\end{equation}
and so
\begin{equation}
    t_{k,n} = \frac{(-1)^n (2n+2k+1) \pi^{2(k+n)}}{(2k+1)k! n! (n+k)!}\ ,
\end{equation}
which can be summed to be:
\begin{equation}
    t_k = \frac{\pi ^k \left(((k-1)!+2 k!) J_k\left(2 \pi \sqrt{E_0} \right)-2 \pi  \sqrt{E_0} (k-1)! J_{k+1}\left(2 \pi \sqrt{E_0} \right)\right)}{(2 k+1) E_0^{(k+1)/2} (k-1)! k!}\ .
\end{equation}

\section{Remarks on perturbative and non-perturbative physics}
\label{sec:exploration}
The central player in this paper is the function $u(x)$, a solution of the underlying string equation~(\ref{eq:big-string-equation}), repeated here for convenience:
\begin{equation}
u{\cal R}^2-\frac{\hbar^2}2{\cal R}{\cal R}^{\prime\prime}+\frac{\hbar^2}4({\cal R}^\prime)^2=\widetilde{\Gamma}^2\ , \quad{\rm with }\quad{\cal R}{\equiv}\sum_{k=1}^\infty t_k R_k[u]{+}x\ ,
\end{equation}
where the $R_k[u]$, polynomials in $u$ and its $x$-derivatives,  were described below~(\ref{eq:big-string-equation}).

Recall from the review material in Section~\ref{sec:introduction} that knowing $u(x)$  unlocks   the full underlying orthogonal polynomial description of the matrix model (in the double scaling limit where gravity emerges) from which all answers about the model can be answered. There is an perturbative expansion of $u(x)$ in $\hbar$, which corresponds to a topological expansion, as well as non-perturbative pieces:
\begin{equation}
    u(x,\hbar) = \sum_{n=0} \hbar^n u_n(x) + \cdots \ ,
\end{equation}
Our analysis has concerned the leading order part, $u_0(x)$, whose allowed (by the string equation) forms determines the allowed features of the leading (disc order) spectral density $\rho_0(E)$. This was(very successfully) compared to various JT supergravity models with extended supersymmetry. It is natural to ask about higher order in perturbation theory, and indeed also about the non-perturbative effects afforded by the formalism, present in fully non-perturbative solutions of the above string equation for~$u(x)$. Below, we will make some remarks about those two topics separately.

\subsection{Perturbative Physics}
\label{sec:perturbative-remarks}
A striking phenomenon that occurred in the formalism was the fact that $u_0(x)$ (starting with the $E_0>\frac14$ case in ${\cal N}{=}2$) fails to be single valued. Such a loss of single-valuedness is already known to be a warning sign~\cite{Johnson:2020lns,Johnson:2021tnl} about the possible lack of a non-perturbative completion. This will be discussed below.

One might worry however that even perturbation theory is affected by the lack of single-valuedness but it is in fact just fine, to all orders. Expanding $u(x)$ in powers of $\hbar$, each term $u_{n+1}(x)$ is determined algebraically by $u_0, u_1,...,u_n$ and a finite number of their derivatives. As such, for $u>E_0$ (which is all that appears in the integrals defining perturbative computations), the multivaluedness does not appear---we established that it always lies at $u_0<E_0$! See {\it e.g.} figures~\ref{fig:u-plots-1} and~\ref{fig:u-plots-2}.

More explcitly, since  we are not interested in the function $u$ for its own sake but rather to determine the spectrum, {\it etc.,} recall that this is doneusing the ``macroscopic loop" operator~(\ref{eq:basic-loop}), repeated here:
\begin{equation}
   Z(\beta)\equiv \langle {\rm Tr}(\e^{-\beta H})\rangle=\int_{-\infty}^\mu\! \langle x|\e^{-\beta{\cal H}}|x\rangle\, dx\ ,\qquad
{\rm where}\qquad
{\cal H} = - \hbar^2 \frac{\partial^2}{\partial x^2} +u(x)\ .
\end{equation}
Since $\cal{H}$ appears in the exponent and contains $x$-derivatives, while in general it would require knowledge of $u$ everywhere, not only for $x\leq\mu$.  performing the small $\hbar$ expansion,
at each order 
we will only encounter values of $u$ for $x \in (-\infty, \mu)$. Thus, $Z(\beta)$ can be safely computed to arbitrary order in $\hbar$. The same kinds of integral appear for higher-point correlators of $Z(\beta)$.

\subsection{Nonperturbative Physics}
\label{sec:non-perturbative-remarks}
Having discussed the perturbative physics (and its agnosticism regarding  multivaluedness), let's turn to the non-perturbative regime. It is clear that if $x(u_0)$ is not surjective, then $u_0 (x)$ cannot be a single-valued function and this is exactly what we find for almost all sectors (i.e., outside the BPS domain). If $u_0 (x)$ is not a single-valued function, then clearly there cannot exist a solution to the full string equation (let us denote it $u(x,\hbar)$) such that:
\begin{equation}
    \lim_{\hbar\to 0} u(x,\hbar) = u_0 (x)\ .
\end{equation}
Thus, the theory does not have a non-perturbative completion, at least as a matrix model of the class considered in this paper.

We saw that there are two sources of multi-valuedness: either minus sign in front of $|\widetilde{\Gamma}|$ (this always leads to non-multivaluedness) or a conspiracy between $t_k$s, such that $u_0(x)$ is not well-defined even with the plus sign in front $|\widetilde{\Gamma}|$. To our best knowledge, the latter type of non-perturbative instability was not encountered before in a supersymmetric setting (see ref.~\cite{Johnson:2020lns} for examples coming from certain models corresponding to deformations of ordinary JT gravity). As such, it deserves further investigation that we postpone to future work.

The instability associated with the minus sign on the other hand was known before. In the context of JT supergravity gravity, it was pointed out in ref.~\cite{Turiaci:2023jfa}, where it was connected to the failure of the original matrix integral to converge. To be more precise, having $\widetilde{\Gamma}$ corresponds to adding logarithmic terms to the matrix potential (as described in section~\ref{sec:introduction}) with the sign determined by the same $\pm$ choice. Thus, the usual matrix model integrand has a factor $\prod_i \lambda_i^{\pm |\Gamma|}$,
where $\lambda_i$ the $i$th eigenvalue of the matrix at hand. If one chooses minus sign, this is a divergence at zero which may make the matrix integral infinite.

Interestingly, solutions with the minus sign choice were found before for multicritical  models at finite $\hbar$ \cite{Carlisle:2005mk}. The construction relied on powerful solution-generating techniques. However, all these techniques failed for $\Gamma \ge 1$ (and minus sign). One can immediately see that this is exactly the threshold when the matrix integral fails to be convergent. Thus, these two observations arrived on the same conclusion albeit from two very different perspectives.

In our work, we actually found the value of $\widetilde{\Gamma} {=} \hbar \Gamma$. Since at $\hbar {=} 0$, $\widetilde{\Gamma}$ is finite it follows that~$\Gamma$ is, in fact, infinite (and, in particular larger than $1$). Thus, for the minus sign, the original matrix integral is definitely ill-defined, in perfect agreement with \cite{Turiaci:2023jfa}. What happens if we increase $\hbar$? Since we keep~$\widetilde{\Gamma}$ fixed, $\Gamma$ must necessarily decrease. That means that there is a certain $\hbar_{\rm min}$ above which $\Gamma <1$ and the matrix integral is convergent. Solutions corresponding to this regime for $\mathcal{N}{=}2$ superJT were constructed in \cite{Johnson:2023ofr} and they seem to be perfectly regular. We are thus led to quite an interesting phase diagram for the theory at hand. Around $\hbar {=} 0$, the theory makes sense to all orders in perturbative expansion but fails at any finite, small $\hbar$. On the other hand, it can be well-defined at $\hbar{>} \hbar_{\rm min}$ (or, in JT gravity language, for sufficiently small $S_0$). It would be interesting to see if there is either a gravitational or SYK interpretation of this phase diagram and to what extent we can extract useful black hole physics from large $\hbar$ regime.

\section{Closing Remarks}
\label{sec:Closing-Remarks}

The primary motivation for this work was to see  to what extent {\it all} JT supergravity models can be recast as multicritical random matrix models. Was the ``miracle'' of ref.~\cite{Johnson:2023ofr} for ${\cal N}{=}2$ (happening again, strikingly,  in ref.~\cite{Johnson:2024tgg} for (small) ${\cal N}{=}4$) a happy accident, or is something deeper going on? Being able to capture the supergravity in multicritical matrix model terms unleashes an extremely efficient methodology for studying many aspects  of the theory (such as correlation functions at arbitrary genus), and even capturing and characterizing non-perturbative physics (giving full computational access to quantities such as the spectral form factor), as was done for the ${\cal N}{=}1$ cases in refs.~\cite{Johnson:2020heh,Johnson:2020exp,Johnson:2021owr,Johnson:2022wsr}.

The results of this paper give a great deal of encouragement to the idea that there is indeed a broad and robust correspondence between  JT supergravity and multicritical models. The rather intricate large ${\cal N}{=}4$ case recently uncovered in ref.~\cite{Heydeman:2025vcc} was found to be fully amenable to a natural multicritical description, and also one of the ${\cal N}{=}3$ cases. Two of the ${\cal N}{=}3$ cases were not as compelling a fit with  our methods, but this may be related to the fact that the underlying supersymmetry algebra has a non-linear component that make them rather different from all the other cases. It remains to be seen whether a different random matrix model framework (perhaps some variant of a 0B theory, as discussed at the end of Section~\ref{sec:Neq3-JT-supergravity}) could work better  for them.

The key question is to what extent a multicritical description can handle the growing power (as~${\cal N}$ grows) of $E$ that can appear in the denominator of the leading spectral density. As already noted in footnote~\ref{fn:pattern}, the naive pattern is that  the spectral density has a $\sinh(\sqrt{E})$ (or $\cosh(\sqrt{E})$) when ${\cal N}$ is odd (or even), and there are ${\cal N}/2$ powers of $E$ in the denominator. That power grows because (roughly speaking) the  R-symmetry group grows in size, and  it becomes increasingly hard to accommodate.  A hard-edge matrix model can really only manage $E^{-\frac12}$, generically, possibly accompanied (see Section~\ref{sec:new-approach}) by a single $E^{-1}$ term for BPS sectors, where the resulting pole's residue  measures how many BPS states are present. These features are handled very straightforwardly in the ${\cal N}{=}1$ and ${\cal N}{=}2$ cases, but what emerges with higher ${\cal N}$ is remarkable. In each case, the peculiarities of the model allowed for ${\cal N}/2$ of~$E$ in the denominator, but in special ways that avoided falling outside the matrix model picture. In the case of small ${\cal N}{=}4$, the potentially dangerous double pole simply goes away for precisely the R-charges allowed. 
Already it is remarkable that this occurs precisely to allow a  matrix model description!

Meanwhile for large ${\cal N}{=}4$, and the ${\cal N}{=}3$ case we studied,  the potential $E^{-{\cal N}/2}$ is split apart into components with smaller powers. This allowed for a splitting of the total density into sums of spectral densities that separately allow a multicritical description, leading us to conjecture (certainly for ${\cal N}{=}4$, more tentatively for ${\cal N}{=}3$) a decomposition into pieces that are statistically independent in the ensemble description. As we noted in the body of the paper, it  would be interesting to see if this prediction of statistical independence can be verified in (for example) a cylinder computation in supergravity.

Very striking, and worth exploring even more (analytically and numerically) is the non-perturbative physics that the full string equation naturally gives access to. (Some of it was discussed in Section~\ref{sec:non-perturbative-remarks}.) For example,  it readily seems to firmly disallow sectors without BPS states that have  non-zero gaps. While these appear  perfectly fine to all orders in perturbation theory, they are non-perturbatively problematic, a result that is very much in line with the general arguments in ref.~\cite{Johnson:2024tgg} connecting BPS states and supergravity gaps. 

As we've seen, many remarkable features of extended supergravity emerged just from analysis of  the leading order string equation. Perhaps there is a great deal more to be learned by further exploring the non-perturbative features.

\section*{Acknowledgements}

We thank Joaquin Turiaci for helpful discussions and comments. MK was supported in part by NSF grant PHY-2408110 and by funds from the University of California. CVJ's  was supported by US Department of Energy grant  \#DE-SC 0011687. Some of CVJ's  work on this  project was done during the Summers of 2024 and 2025 at the Aspen Center for Physics, which is supported by National Science
Foundation grant PHY-2210452. CVJ also thanks Amelia for her support and patience.


\bigskip
\bigskip
\bigskip

\appendix

\noindent 
{\bf \Large Appendices}

\section{A universal formula for computing $t_k$s}
\label{app:teekay-machinery}
Let us define the following  polynomial of degree $k{-}1$:
\begin{eqnarray}
    F_k(\lambda)= {}_2 F_1 \left(1-k, 1, \frac{3}{2},-\lambda \right)\ .
\end{eqnarray}
 We wish to find a family of (infinite order) operators $O_j$ such that:
\begin{equation}
    O_j F_k(\lambda = 0) = \delta_{jk}\ .
\end{equation}
If we have them, we can read off $t_k$s from the action of $O_j$ on the density function:
\begin{equation}
    \tilde{\rho}(\lambda, E_0) = \lambda^{-1/2} \rho_0(\lambda, E_0) \mp \frac{|\widetilde\Gamma|}{2\pi \hbar} \frac{1}{E_0 (1+\lambda)}\ .
\end{equation}
Note that $\tilde\rho(\lambda)$ is a smooth function. It will be convenient to introduce the following coefficients:
\begin{equation}
    a_n = \partial_\lambda^n F_{n+1}(\lambda=0) = \frac{2^n (n!)^2}{(2n+1)!!} = \frac{2^{2n} (n!)^3}{(2n+1)!}\ .
\end{equation}
Then, it is easy to notice that the following works:
\begin{eqnarray}
    O_1 = \sum_{n=0} \frac{(-1)^n}{a_n} \partial_\lambda^{n}\ .
\end{eqnarray}
 What about higher $O$s? The following do the job:
\begin{equation}
    O_2 = \sum_{n=1} \frac{(-1)^{n+1} n}{a_n}\partial^n_\lambda\ ,
\end{equation}
\begin{equation}
    O_3 = \frac{1}{2} \sum_{n=2} \frac{(-1)^n n(n-1)}{a_n} \partial_\lambda^n\ ,
\end{equation}
\begin{equation}
    O_4 =\frac{1}{6} \sum_{n=3} \frac{(-1)^{n+1} n(n-1)(n-2)}{a_n} \partial_\lambda^n\ ,
\end{equation}
\begin{eqnarray}
    O_5=\frac{1}{24}  \sum_{n=4} \frac{(-1)^n n(n-1)(n-2)(n-3)}{a_n} \partial_\lambda^n\ .
\end{eqnarray}
We thus see a clear pattern and we can write:
\begin{equation}
    O_k = \sum_{n=k-1} \frac{(-1)^{n+k-1}}{a_n} {{n}\choose{k-1}} \partial_\lambda^n\ .
\end{equation}
This can be written even more compactly if we introduce: 
\begin{equation}
    O(y) =  \sum_{n=0} \frac{(y-1)^n}{a_n} \partial_\lambda^{n}\ .
\end{equation}
Then, it is easy to see that:
\begin{equation}
    O_k = \frac{1}{(k-1)!}\partial_y^{k-1} O(y=0)\ ,
\end{equation}
or
\begin{equation}
    O(y) = \sum_{k=0} \frac{y^k}{k!} \partial_y^{k} O(y=0) = t^k O_{k+1}\ ,
\end{equation}
and so it follows that
\begin{equation}
    O(y) \hat{\rho}(\lambda = 0)
\end{equation}
is a generating function of the $t_k$s.

\bibliographystyle{utphys}
\bibliography{Refs,super_JT_gravity1,super_JT_gravity2,extra_references,Fredholm_super_JT_gravity1,Fredholm_super_JT_gravity2}
\end{document}